\documentclass{ws-ijmpe}
\usepackage[super,compress]{cite}
\usepackage{color}
\usepackage{pdflscape}
\usepackage[utf8]{inputenc}
\usepackage[english]{babel}
\usepackage{hyperref}
\begin{document}
\markboth{A. Jain \textit{et al.}}{Theoretical Investigation of $\alpha$-decay in Heavy and Superheavy Isomers}
\catchline{}{}{}{}{}
\title{Theoretical Investigation of $\alpha$-decay in Heavy and Superheavy Isomers}
\author{A. Jain}
\address{Department of Physics, School of Physical and Biosciences, Manipal University Jaipur, Jaipur-303007, India}
\author{M. Imran}
\address{{Applied Science \& Humanities Section, University Women’s Polytechnic, Aligarh Muslim University, Aligarh-202002, India},\\
{Department of Physics, Aligarh Muslim University, Aligarh-202002, India}}
\author{G. Saxena}
\address{Department of Physics (H\&S), Govt. Women Engineering College, Ajmer-305002, India\\ gauravphy@gmail.com}

\maketitle

\begin{history}
\received{Day Month Year}
\revised{Day Month Year}
\end{history}
\begin{abstract}
The heavy and superheavy elements of the periodic table predominately disintegrate by $\alpha$-decay, facilitating transitions mainly between ground states and occasionally involving isomeric states. This study focuses on estimating the half-lives of $\alpha$-transitions both from and to isomeric states, using a recently refined formula which shows excellent agreement with experimental data when isospin of parent nucleus as well as angular momentum taken away by the $\alpha$ particle are incorporated. These findings provide valuable insights for upcoming experimental investigations of isomeric states. Additionally, the study predicts potential $\alpha$-decay in several yet-unobserved isomeric nuclei, contributing to a deeper understanding of nuclear structure in heavy and superheavy elements.
\keywords{Isomer, $\alpha$-decay; Half-life; Decay Chains; Heavy Nuclei; Superheavy Nuclei.}
\end{abstract}

\section{Introduction}
Several nuclear decay modes have been experimentally observed, such as \( \alpha \)-decay, \( \beta \)-decay, proton emission, cluster decay, and spontaneous fission. Among these, $\alpha$-decay is particularly significant for heavy and superheavy nuclei, offering critical insights into nuclear structure \cite{oganessian2010synthesis,Og2006,Wuenschel2018,audi20201,Voinov2020,heinz2012,oganessian2011eleven,og2015npa}. The process of $\alpha$-decay serves as a crucial tool for discovering new isotopes and elements, as detecting this decay enables the identification of an unknown parent nucleus through its transition to a well-characterized daughter nucleus. Initially detected by Rutherford and later interpreted by Gamow in 1928 as a quantum tunneling effect \cite{gamow1928quantentheorie}, $\alpha$-decay continues to be a fundamental aspect of nuclear physics. It reveals valuable details about nuclear properties such as ground-state energy, half-life, clustering, shell effects, deformation, spin, parity, and nuclear interactions \cite{hofmann2000discovery,hamilton2013search,heenen2015shapes,oganessian2015super,ni2013}. This decay phenomenon predominantly manifests within heavy, superheavy, and neutron-deficient nuclei characterized by a significantly elevated protons-to-neutron ratio.\par

Recent experimental breakthroughs \cite{wang2024decay} and the synthesis of eleven new superheavy nuclei \cite{oganessian2010synthesis}, have significantly advanced our understanding of nuclear structure. Theoretical frameworks based on quantum tunneling models provide crucial insights into
$\alpha$-decay \cite{razavy2013quantum}, while numerous theoretical studies have explored
$\alpha$-decay transitions, particularly focusing on ground-state decays \cite{manjunatha2019, newrenA2019, horoi2004, renA2004, qi2009, parkho2005, taageper1961relations, qian2012unfavored, singh2020, Saxena2021jpg, sharma2021npa, budaca2016, poenaru2007, royer2010, MYQZR2019, budaca2021screening, akrawy2022generalization}. Nonetheless, our understanding of the structural characteristics of isomeric states in many nuclei is still limited, and there is a notable lack of research focused on structure calculations.\par

Recent strides in experimental research on isomers have reasonably enriched our understanding of nuclear structure. Jain \textit{et al.} \cite{jain2021nuclear} provided a comprehensive compilation of isomers across the periodic chart, further expanded upon in their subsequent work \cite{garg2023atlas}. Arnquist \textit{et al.} \cite{arnquist2023constraints} pushed the boundaries of isomeric decay studies by employing an unprecedented 17 kilograms of tantalum metal to investigate the rare nuclear heavy isomer $^{180}$Ta$^{m}$. Their groundbreaking research substantially refined existing estimates of the $^{180}$Ta$^{m}$ half-life, enhancing precision by one to two orders of magnitude. In addition to these advancements, recent experimental endeavors have uncovered notable superheavy isomers such as $^{258}$Rf$^{m,n}$ and $^{257}$Lr$^{m}$ \cite{Antalic2016}, among others. While many of these isomers, decay through internal transitions or gamma-ray emission, as documented in NUBASE2020 \cite{audi20201} and also shown on NNDC database \cite{nndc}, certain cases exhibit a preference for $\alpha$-decay. Particularly, in heavy nuclei like $^{183}$Pb$^{m}$ (a magic nucleus) and $^{186}$Bi$^{m}$, $\alpha$-decay emerges as a prominent decay mode. Similarly, superheavy nuclei such as $^{259}$Sg$^{m}$ and $^{257}$Rf$^{m}$ predominantly decay via $\alpha$-emission, with nearly 100$\%$ probability. These findings highlight the diverse decay mechanisms operating within nuclei across the atomic landscape. \par
Theoretical explorations have underscored the importance of isomeric states in bolstering the stability of nuclei, often exhibiting lifetimes that surpass those of ground states \cite{Xu2004,Herzberg2006}. To gain deeper insights into this phenomenon, theoretical studies have been conducted on $\alpha$-decay transitions from isomeric states to either ground states or other isomeric states across a wide range of nuclei \cite{Chang2007,Chang2006,Delion2006,Peltonen2008,Zhangen2008,Wang2009,poenaru1981alpha,ni2010alpha,santhosh2010alpha,qian2011alpha,sun2017systematic,sun2017}. Noteworthy among these are the studies by Denisov and Khudenko \cite{Denisov2009} and Dong \textit{et al.} \cite{Dong2010}, which examined $\alpha$-decay to ground and isomeric states of heavy deformed nuclei employing unified models and the GLDM (generalized liquid drop model), respectively. More recently, Clark's investigation \cite{clark2024hindrances} delved into the hindrances to $\alpha$-decay and fission processes in isomers. Anjali \textit{et al.} conducted a detailed study of the $\alpha$-decay properties stemming from both ground and isomeric states across nuclei within the range of 67 $\leq$ Z $\leq$ 106 \cite{Anjali2024Systematic}. Collectively, these investigations serve as compelling motivation for our current study.\par

Isomers characterized by high spin states typically exhibit $\alpha$-decay with elevated angular momentum ($l$), a feature discerned through selection rules \cite{denisov2009alpha}. This trend is evident in recent discoveries such as the high spin (19$^{-}$) isomeric state of $^{156}$Lu reported by Lewis \textit{et al.} in 2018 \cite{PhysRevC.98.024302}, as well as the spin (39/2)$^{-}$ observed for a high-spin isomer in $^{93}$Mo \cite{fukuchi2005}. The angular momentum ($l$) plays a pivotal role in determining the half-lives of $\alpha$-decay for these isomers. Our objective is to identify the formula that best accommodates this angular momentum ($l$) term, recognizing its significance in characterizing the decay process. \par


For accurate prediction of $\alpha$-decay half-lives, the angular momentum ($l$) carried away by the emitted $\alpha$ particle is a critical factor. This aspect has been thoroughly addressed in several recently refined formulas developed for ground-to-ground state $\alpha$-transitions \cite{NMTN,ismail2022improved,saxena2021,saxena2023cluster,newrenA2019,sharma2021npa,royer2020,soylu2021,akrawy2018,akrawy2018new,akrawy2022generalization,cheng2022isospin,MYQZR2019,denisov2010decay,Akrawymrf2018}. The present study does not propose yet another new formula; instead, it aims to evaluate and compare the predictive performance of existing empirical models using various statistical indicators, applied uniformly to a consistent dataset of isomeric $\alpha$-transitions \cite{audi20201,audi20202,nndc}. Based on this comparative analysis, the formula demonstrating the best overall agreement with known data is used to predict the logarithmic half-lives of $\alpha$-decays originating from isomeric states in 23 nuclei, as listed in NUBASE2020 \cite{audi20201} and the NNDC database \cite{nndc}. Furthermore, potential $\alpha$-transitions involving isomeric states are investigated. This comprehensive approach has led to the identification of several promising $\alpha$-decay chains, some of which involve isomeric levels, thereby contributing to the potential discovery of new isomeric states and deepening our understanding of $\alpha$-decay mechanisms, especially in the heavy and superheavy nuclear regions.\par
\section{Formalism}
Isomers encompass a spectrum of energy levels known as isomeric states, denoted by symbols like (m), (n), and beyond. When an isomer undergoes $\alpha$-emission, the resulting daughter nucleus may either reside in its ground state (with 0 MeV energy) or any isomeric state. Likewise, a parent nucleus in its ground state can decay into any isomeric state of the daughter nucleus. These transitions—namely, from isomeric to ground, isomeric to isomeric, and ground to isomeric—are commonly termed isomeric $\alpha$-transitions. Interestingly, even when the neutron number, proton number, and mass number remain identical for both parent and daughter nuclei, the half-lives of these transitions differ. This disparity arises primarily from the $Q_{\alpha}$-value, which signifies the energy liberated in $\alpha$-decay and serves as a key determinant of half-lives. While in ground-to-ground state $\alpha$-transitions, the $Q_{\alpha}$ value is typically computed using the B.E. (binding energies) of the daughter, parent nuclei, and $\alpha$-particle, in isomeric transitions, it hinges on the E$_{id}$ (excitation energy of the daughter nucleus) and the E$_{jp}$ (excitation energy of the parent nucleus) \cite{Denisov2009}. This calculation is as follows:
\begin{equation}\label{Q}
Q_{j{\rightarrow}i} = Q_{g.s.{\rightarrow}g.s.}-E_{id}+E_{jp}
\end{equation}
where $Q_{g.s.{\rightarrow}g.s.}$ is the disintegration energy of $\alpha$-decay released in ground-to-ground state. \par

In 1911, Geiger and Nuttall identified a correlation between $\alpha$-decay half-lives and $Q_{\alpha}$-values by introducing a straightforward linear relation ($\log_{10}T_{1/2} = aQ^{-1/2} + b$) \cite{geiger1911}, which later became known as the Geiger-Nuttall law. However, in 2000, Royer generalized this law \cite{royer2000} due to its dependence on two adjustable parameters ($a$ and $b$), which vary among different $\alpha$-decay radioactive isotopes. This variability renders the Geiger-Nuttall law challenging to universally apply across all $\alpha$-decay cases and limits its predictive power. According to Royer, half-lives are proportional to $Z/\sqrt{Q_{\alpha}}$, where $Z$ denotes the atomic number of the parent nucleus. Subsequently, numerous modifications to this law have been proposed by various authors \cite{manjunatha2019,newrenA2019,horoi2004,renA2004,qi2009,parkho2005,taageper1961relations,qian2012unfavored,singh2020,Saxena2021jpg,sharma2021npa,budaca2016,poenaru2007,royer2010,MYQZR2019,budaca2021screening,akrawy2022generalization}.
For estimating the half-lives of $\alpha$-decay from isomeric transitions, a straightforward approach is to employ such empirical formulas primarily reliant on $Q_{\alpha}$-values. However, in the context of isomers, the angular momentum ($l$) often assumes high values for isomeric $\alpha$-transitions. For instance, consider the case of $^{211}$Po$^{m}$ decaying into $^{207}$Pb, where the $\alpha$-particle carries $l$ value of 13. Given that angular momentum ($l$) significantly influences $\alpha$-decay half-lives, as elucidated by various authors \cite{Chang2007,Chang2006,Delion2006,Peltonen2008,Zhangen2008,Wang2009,Denisov2009,Dong2010}, it stands to reason that empirical formulas incorporating $l$-dependent terms offer more accurate predictions. In nutshell, when striving for precise estimations of $\alpha$-decay half-lives from isomeric transitions, empirical formulas equipped to accommodate angular momentum ($l$) dependencies \cite{ismail2022improved,saxena2021, saxena2023cluster,newrenA2019,sharma2021npa,royer2020,soylu2021,akrawy2018,akrawy2018new,akrawy2022generalization,cheng2022isospin,MYQZR2019,denisov2010decay,Akrawymrf2018} present a preferable choice. The calculation of $l$ follows the selection rules outlined in \cite{denisov2009alpha}:
\begin{equation}
   l=\left\{
    \begin{array}{ll}
       \triangle_j\,\,\,\,\,\,
       &\mbox{for even}\,\,\triangle_j\,\,\mbox{and}\,\,\pi_{i} = \pi_{f}\\
       \triangle_{j}+1\,\,\,\,\,\,
       &\mbox{for even}\,\,\triangle_j\,\,\mbox{and}\,\,\pi_{i} \neq \pi_{f}\\
       \triangle_{j}\,\,\,\,\,\,
       &\mbox{for odd}\,\,\triangle_j\,\,\mbox{and}\,\,\pi_{i} \neq \pi_{f}\\
       \triangle_{j}+1\,\,\,\,\,\,
       &\mbox{for odd}\,\,\triangle_j\,\,\mbox{and}\,\,\pi_{i} = \pi_{f}\\
      \end{array}\right.
      \label{lmin}
\end{equation}
where $\triangle_j = |j_p - j_d|$, with $j_p$, $\pi_p$, $j_d$, and $\pi_d$ representing the spin and parity values of the parent and daughter nuclei, respectively. In this study, these values are primarily taken from the latest evaluated nuclear properties table, NUBASE2020 \cite{audi20201}, and the NNDC database \cite{nndc}. If unavailable, values are adopted from Ref. \cite{moller2019}.\par

Apart from the influence of $l$, a crucial factor that can significantly impact the decay properties of nuclei is the isospin of the parent nucleus, given by $I = (N - Z)/A$. Several authors \cite{sharma2021npa,singh2020,soylu2021} have noted the considerable sensitivity of $\alpha$-decay half-life to this factor. In our study, which focuses on the heavy and superheavy region where isospin asymmetry is more pronounced, the reliance on isospin asymmetry becomes essential. Recently, Seif \textit{et al.} emphasized the significant role of this factor ($I$) in isomeric $\alpha$-transitions \cite{seif2023stability}. \par



Several empirical formulas incorporating microscopic effects—such as angular momentum ($l$) and isospin asymmetry ($I$)—have been analyzed, as these factors significantly affect energy levels, barrier penetration probabilities, and nuclear stability. Consequently, they play a crucial role in determining the $\alpha$-decay half-lives, particularly in isomeric states. To identify the most suitable expression for predicting $\alpha$-decay half-lives, various recent $l$-dependent formulas have been examined, including the New Modified Horoi Formula(NMHF) \cite{sharma2021npa}, MYQZR \cite{MYQZR2019}, Improved Unified Formula (IUF) \cite{ismail2022improved}, Modified RenB Formula (MRenB) \cite{newrenA2019}, New Modified Sobiczewski Formula (NMSF) \cite{sharma2021npa}, Royer 2020 Formula \cite{royer2020}, Modified Taagepera and Nurmia Formula (MTNF) \cite{saxena2021}, Modified Budaca Formula (Mbudaca) \cite{akrawy2022generalization}, Quadratic Fitting Formula (QF) \cite{saxena2021}, New Modified Taagepera and Nurmia Formula (NMTN) \cite{NMTN}, MUDL \cite{soylu2021}, Denisov-Khudenko Formula (DK1) \cite{denisov2010decay}, Akrawy Formula \cite{akrawy2018new}, Improved Tavares and Medeiros Formula (ITM) \cite{saxena2023cluster}, New Modified Manjunatha Formula (NMMF) \cite{sharma2021npa}, Denisov-Khudenko Formula (DK2) \cite{akrawy2018}, Modified Royer Formula (MRF) \cite{Akrawymrf2018}, and Improved Semi-Empirical Formula (ISEF) \cite{cheng2022isospin}.\par
It is noteworthy that all 18 of these models were originally parametrized using ground-state-to-ground-state ($g.s. \rightarrow g.s.$) $\alpha$-decay data. For a more rigorous and unbiased comparison in the context of isomeric states, each formula must be re-parametrized using exclusively the isomeric $\alpha$-decay dataset \cite{audi20201,audi20202,nndc}, which includes 165 transitions in the present analysis. Accordingly, these formulas have been re-fitted using this dataset. For the convenience of the reader, all 18 formulas along with their newly determined coefficients are compiled in Appendix A.\par
For the comparison, we use few statistical parameters such as root mean square error(RMSE), $\chi ^2$ per degree of freedom ($\chi ^2$), standard deviation ($\sigma$), uncertainty ($u$), average deviation factor ($\overline{x}$), and mean deviation $\overline{\delta}$. All these statistical parameters for these formulas are mentioned in Table \ref{RMSE_table}. These statistical parameters are defined as:
\begin{equation}
\text{RMSE}  = \sqrt{\frac{1}{N_{nucl}}\sum^{N_{nucl}}_{i=1}\left(log\frac{T^i_{Th.}}{T^i_{Exp.}}\right)^2}
\label{rmse}
\end{equation}
\begin{equation}
\chi^2 = \frac{1}{N_{nucl}-N_{p}}\sum^{N_{nucl}}_{i=1}\left(log\frac{T^i_{Th.}}{T^i_{Exp.}}\right)^2
\label{kai}
\end{equation}

\begin{equation}
\sigma = \sqrt{\frac{1}{N_{nucl}-1}\sum^{N_{nucl}}_{i=1}\left(log\frac{T^i_{Th.}}{T^i_{Exp.}}\right)^2}
\end{equation}

\begin{equation}
u =  \sqrt{\frac{1}{N_{nucl}(N_{nucl}-1)}\sum^{N_{nucl}}_{i=1}\left(log\frac{T^i_{Th.}}{T^i_{Exp.}}-\mu \right)^2}
\label{un}
\end{equation}

\begin{equation}
\overline{x} = \frac{1}{N_{nucl}}\sum^{N_{nucl}}_{i=1}\left(\frac{|logT^i_{Exp.}-logT^i_{Th.}|}{logT^i_{Exp.}}\right)
\end{equation}

\begin{equation}
\overline{\delta} = \frac{1}{N_{nucl}}\sum^{N_{nucl}}_{i=1}\left|log\frac{T^i_{Th.}}{T^i_{Exp.}}\right|
\end{equation}
where, $N_{nucl}$ is the total number of nuclei (data) and $N_{p}$ is the number of degree of freedom (or no. of coefficients). $T^i_{Exp.}$ and
$T^i_{Th.}$ are the experimental and theoretical values of half-lives for $i^{th}$ data point, respectively.
\begin{table}[!htbp]
\caption{\label{RMSE_table}Various statistical parameters of different formulas for $\alpha$-decay half-life, incorporating $l$ dependent term(s), utilizing a dataset 165 isomeric $\alpha$-transitions exclusively from pure isomeric transitions.}
\centering
\footnotesize
\begin{tabular}{l@{\hskip 0.4in}c@{\hskip 0.4in}c@{\hskip 0.4in}c@{\hskip 0.4in}c@{\hskip 0.4in}c@{\hskip 0.4in}c}
\hline
Formula&RMSE &$\chi ^2$&$\sigma$& $u$&$\overline{x}$&$\overline{\delta}$\\

\hline
NMHF \cite{sharma2021npa}                &0.6061&0.3765&0.6080&0.0473&0.5190&0.4128 \\
MYQZR \cite{MYQZR2019}                   &0.6104&0.3866&0.6030&0.0469&0.5304&0.4075 \\
IUF \cite{ismail2022improved}            &0.6125&0.3844&0.6143&0.0478&0.5413&0.4183 \\
MRenB \cite{newrenA2019}                 &0.6129&0.3850&0.6148&0.0479&0.5554&0.4197 \\
NMSF \cite{sharma2021npa}                &0.6155&0.3883&0.6135&0.0478&0.3907&0.4208 \\
Royer2020 \cite{royer2020}               &0.6245&0.3997&0.6227&0.0485&0.4402&0.4257 \\
MTNF \cite{saxena2021}                   &0.6384&0.4150&0.6403&0.0498&0.5746&0.4515 \\
Mbudaca \cite{akrawy2022generalization}  &0.6459&0.4276&0.6478&0.0504&0.4205&0.4203 \\
QF \cite{saxena2021}                     &0.6573&0.4428&0.6593&0.0513&0.4528&0.4512 \\
NMTN \cite{NMTN}                         &0.6600&0.4906&0.6720&0.0523&0.5942&0.4496 \\
MUDL \cite{soylu2021}                    &0.6877&0.4846&0.6898&0.0537&0.5104&0.4728 \\
DK1 \cite{denisov2010decay}              &0.7441&0.5674&0.7464&0.0581&0.6797&0.5144 \\
Akrawy \cite{akrawy2018new}              &0.7466&0.5713&0.7489&0.0583&0.4685&0.5164 \\
ITM \cite{saxena2023cluster}             &0.7885&0.6372&0.7909&0.0616&0.4778&0.5699 \\
NMMF \cite{sharma2021npa}                &0.8203&0.6896&0.8228&0.0641&0.6593&0.5694 \\
DK2 \cite{akrawy2018}                    &0.8234&0.6992&0.8259&0.0643&0.5734&0.5434 \\
MRF \cite{Akrawymrf2018}                 &0.8512&0.7426&0.8517&0.0663&0.6702&0.5953 \\
ISEF \cite{cheng2022isospin}             &0.8586&0.7555&0.8612&0.0670&0.7887&0.6065 \\
\hline
\end{tabular}\\
\end{table}
 As evident from Table \ref{RMSE_table}, the NMHF formula exhibits the best performance among all 18 considered models in predicting the half-lives of isomeric $\alpha$-transitions. This formula is a modified version of the well-known Horoi formula \cite{horoi2004}. Originally proposed by M. Horoi in 2004, the semi-empirical Horoi Formula (HF) was developed for cluster decay and has since found wide application in $\alpha$-decay studies due to its strong physical foundation. Notably, Horoi identified a linear dependence between the logarithmic half-life and the scaling variable $S = (Z_{\alpha}Z_d)^{0.6}/\sqrt{Q_\alpha}$, establishing a model-independent relationship that offers deep insight into the underlying decay dynamics.\par
 In the recent work by Saxena et al. \cite{Saxena2021jpg}, the original formula was extended by incorporating the asymmetry dependence, leading to the Modified Horoi Formula (MHF). This updated formulation has been successfully applied to describe $\alpha$-decay chains, accounting for the competition between $\alpha$-decay and spontaneous fission. Moreover, the consideration of angular momentum carried away by the emitted $\alpha$-particle has led to the development of a new variant, termed the New Modified Horoi Formula (NMHF) \cite{sharma2021npa}, which is presented as follows:
 \begin{equation}
\log_{10} T_{1/2}^{\text{NMHF}} (\text{sec}) = (a \sqrt{\mu} + b)\left[(Z_\alpha Z_d)^{0.6} Q_\alpha^{-1/2} - 7\right] + (c \sqrt{\mu} + d) + eI + fI^2 + g l(l + 1)
\end{equation}
Here, the parameters \( a, b, c, d, e, f, \) and \( g \) are fitting coefficients, which have been re-evaluated using the isomeric \(\alpha\)-decay dataset \cite{audi20201,audi20202,nndc}. The resulting optimized values are \( a = 136.1615 \), \( b = -263.6300 \), \( c = -172.9266 \), \( d = 334.7084  \), \( e = 11.1648  \), \( f = -18.5868 \), and \( g =  0.0409 \), respectively.

\section{Results and Discussions}
To evaluate the influence of centrifugal terms on the $\alpha$-decay half-life of isomeric states, a comparative analysis has been conducted between recently developed $l$-dependent formulas and their original $l$-independent counterparts. The $l$-dependent formulas examined include the NMMF \cite{sharma2021npa}, MUDL \cite{soylu2021}, NMTN \cite{NMTN}, Royer 2020 Formula \cite{royer2020}, NMSF \cite{sharma2021npa}, and NMHF \cite{sharma2021npa}, each specifically adapted for isomeric $\alpha$-transitions through refitting on the present dataset. For a fair and consistent comparison, the corresponding $l$-independent versions of these formulas are also analyzed and re-fitted using the same dataset. These include the original formulations by Manjunatha \textit{et al.} \cite{manjunatha2019}, the Universal Decay Law (UDL) \cite{qi2009}, the Taagepera and Nurmia Formula (TN) \cite{taageper1961relations}, Royer \cite{royer2010}, Sobiczewski \cite{parkho2005}, and Horoi \cite{horoi2004}, respectively.

\begin{figure*}[!htbp]
\centering
\includegraphics[width=0.74\textwidth]{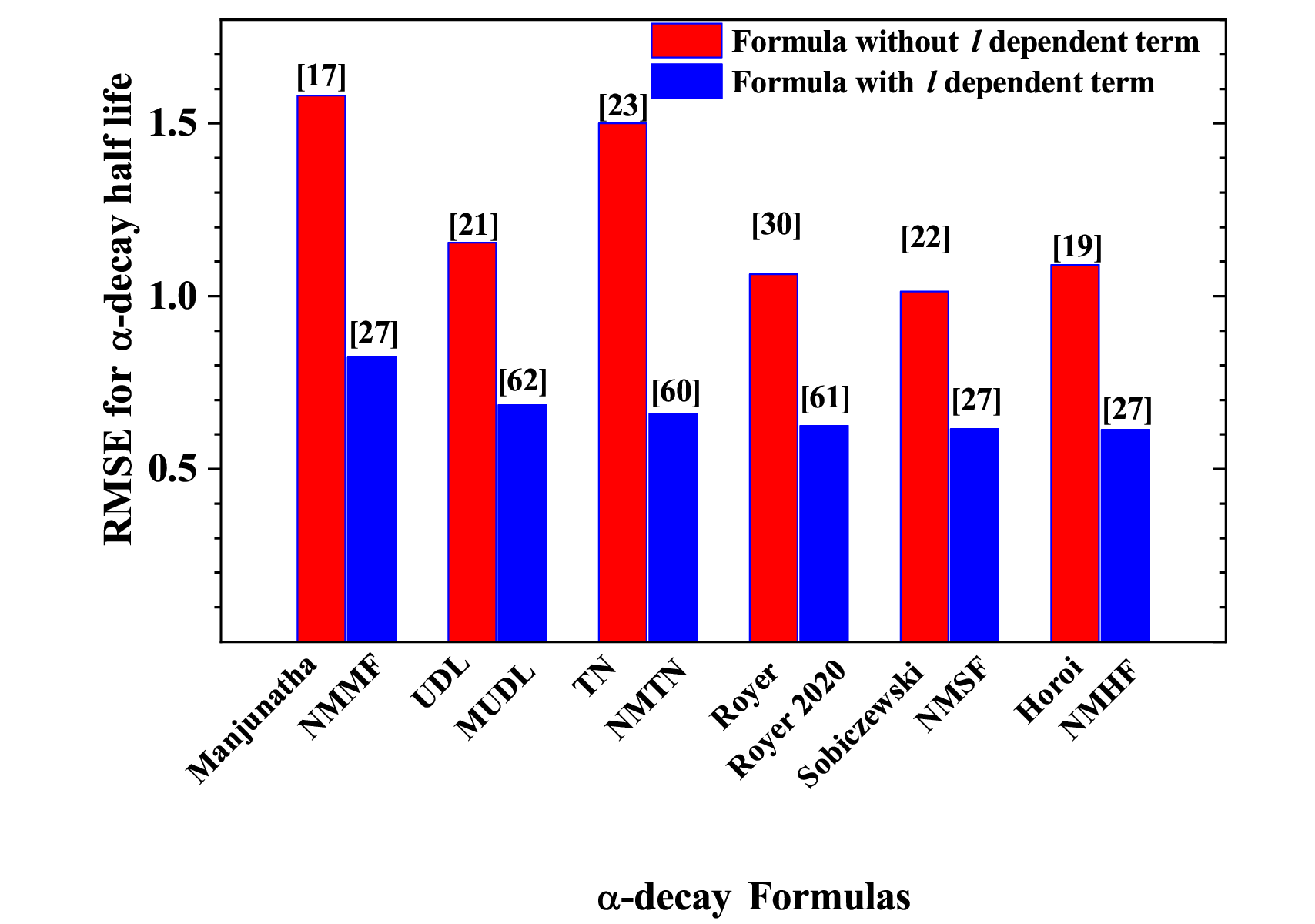}
\caption{(Colour online) Comparison of RMSE between two types of formulas: 1. Formulas without $l$-dependent term(s) (indicated by red bars) 2. Formulas with $l$-dependent term(s) (indicated by blue bars). References for each type of formula are provided above the corresponding bar.}
\label{rmse}
\end{figure*}

The $Q_{\alpha}$ value, a critical factor affecting $\alpha$-decay half-life, is calculated using Eqn. (\ref{Q}), while the angular momentum ($l$) is determined using the selection rules outlined in Eqn. (\ref{lmin}). Using 165 experimental data points of $\alpha$-decay in isomers, we have plotted the root mean square error (RMSE) in Fig. \ref{rmse}. The red bars representing the RMSE of formulas without $l$-dependent term(s) consistently exceeded the blue bars representing the RMSE of formulas with $l$-dependent term(s). This observation underscores the role of $l$-dependent term(s) in reducing RMSE and improving the accuracy of $\alpha$-decay half-life predictions in isomers.\par


\begin{figure*}[!htbp]
\centering
\includegraphics[width=0.92\textwidth]{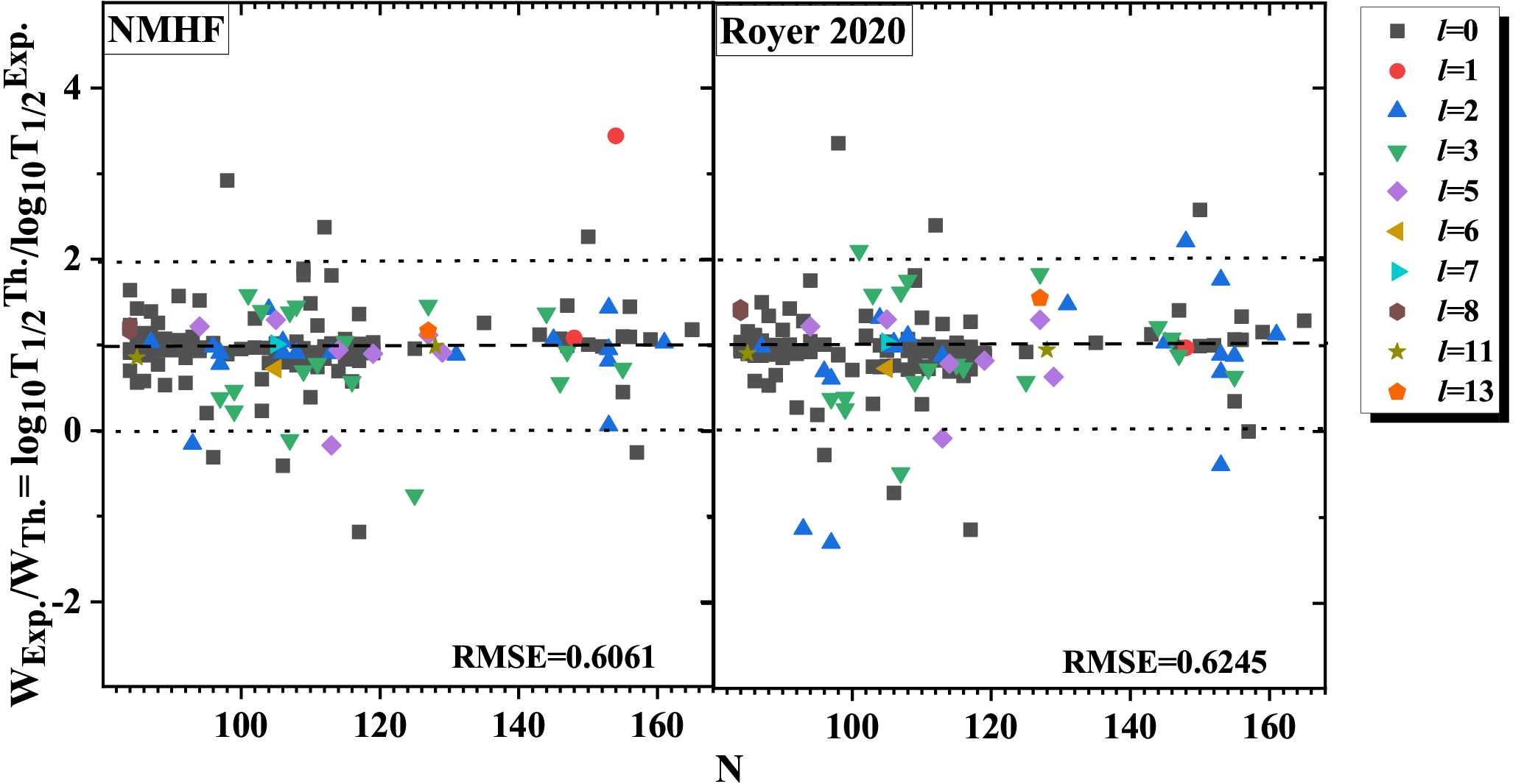}
\caption{Ratio of experimental to theoretical decay widths for 165 isomeric $\alpha$-transitions as a function of the N (neutron number of parent nuclei).}
\label{decay_width}
\end{figure*}

The analysis proceeds with a particular focus on one of the most widely used and recently modified $l$-dependent formulas: the Royer 2020 formula. Alongside the NMHF, both models are employed to predict $\alpha$-decay transitions originating from or occurring within isomeric states of heavy and superheavy nuclei. To the best of current knowledge, the Royer 2020 formula is unique among the considered models in incorporating fittings based not only on ground-state-to-ground-state ($g.s. \rightarrow g.s.$) transitions but also on isomeric $\alpha$-decays. This distinctive feature motivates its selection as a benchmark for fair and meaningful comparison, as demonstrated in Fig. \ref{decay_width}.

The RMSE values obtained for isomeric $\alpha$-transitions are 0.6061 for the NMHF and 0.6245 for the Royer 2020 formula \cite{royer2020}, indicating improved predictive accuracy in our approach. To visualize this, we plot the ratio of experimental to theoretical decay widths ($W_{\text{Exp.}}/W_{\text{Th.}} = \log_{10}T_{1/2}^{\text{Th.}} / \log_{10}T_{1/2}^{\text{Exp.}}$) as a function of neutron number ($N$) in Fig. \ref{decay_width}. This comparison is conducted for a fixed angular momentum transfer $l$. Across all considered values of $l$, our NMHF with present fitting yields more accurate estimations, with decay width ratios closer to unity, demonstrating better agreement with experimental data than the Royer 2020 formula.
\begin{figure*}[!htbp]
\centering
\includegraphics[width=0.7\textwidth]{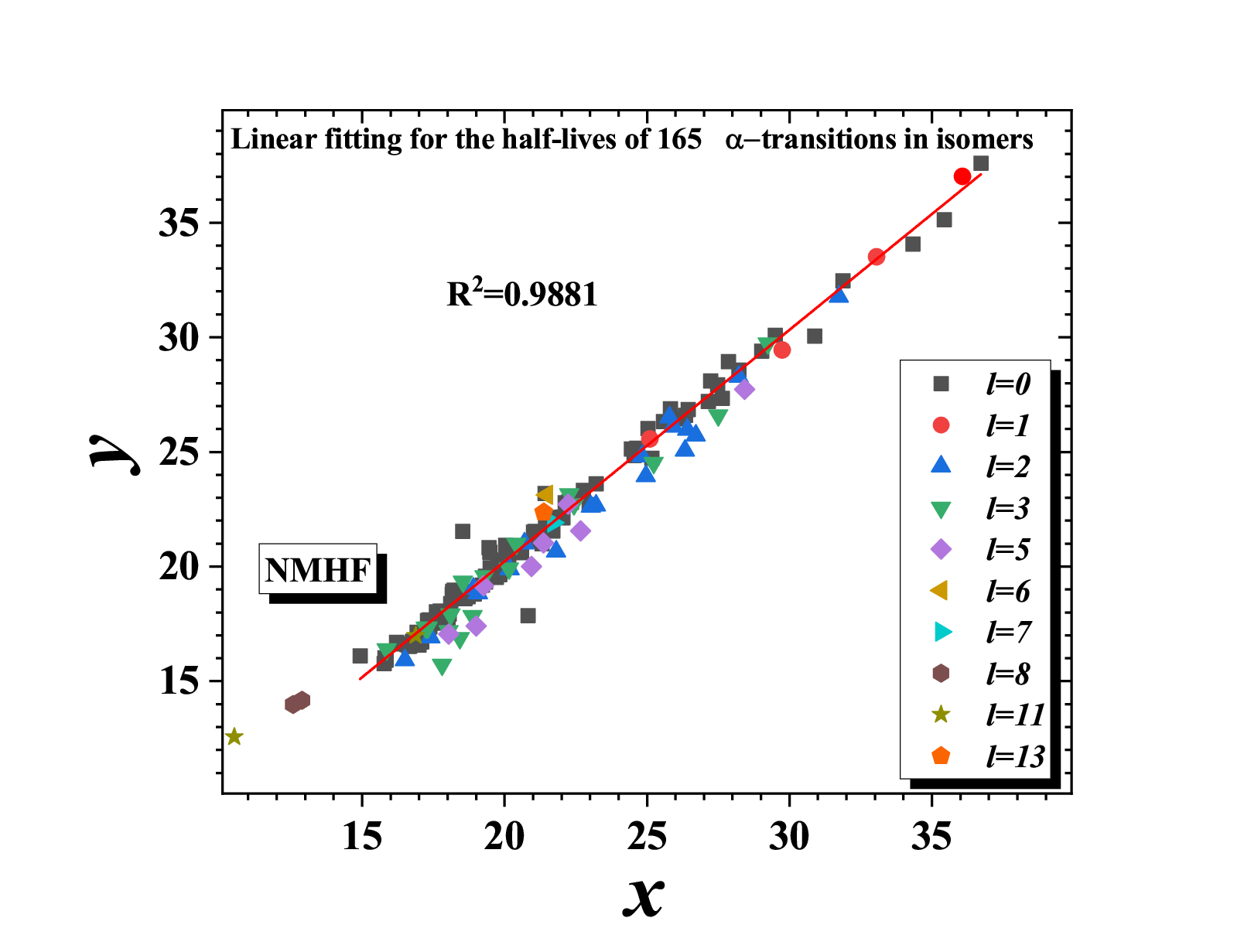}
\caption{Linear fitting using 165 experimental isomeric $\alpha$-transitions in which x-axis shows $x=(a\sqrt{\mu} + b)[(Z_{\alpha}Z_{d})^{0.6}Q_{\alpha}^{-1/2} - 7] + (c\sqrt{\mu} + d) + eI + fI^{2}$ and y-axis is $y=log_{10}T_{1/2}^{Exp}-g l (l + 1)$.}
\label{linear_fittings}
\end{figure*}
However, in addition to the $l$ dependence, the conventional linear dependence on ($1/\sqrt{Q_{\alpha}}$) \cite{geiger1911} should be assured. With this in view, we have plotted Fig. \ref{linear_fittings}, in which the linear fitting is obtained for the NMHF using the same experimental dataset. In this Fig. \ref{linear_fittings}, we have shown the fluctuation in the relation $y$=$log_{10}T_{1/2}^{Exp}-g l (l + 1)$ with the $Q_{\alpha}$ and $I$ dependent terms of NMHF formula with present fitting. The points are positioned near the straight line for all the considered $l$ values which lead to R$^{2}$=0.9881 representing a statistical measure of fit (Usually, a higher R$^{2}$ indicates a better fit for the model).

\begin{figure*}[!htbp]
\centering
\includegraphics[width=1.0\textwidth]{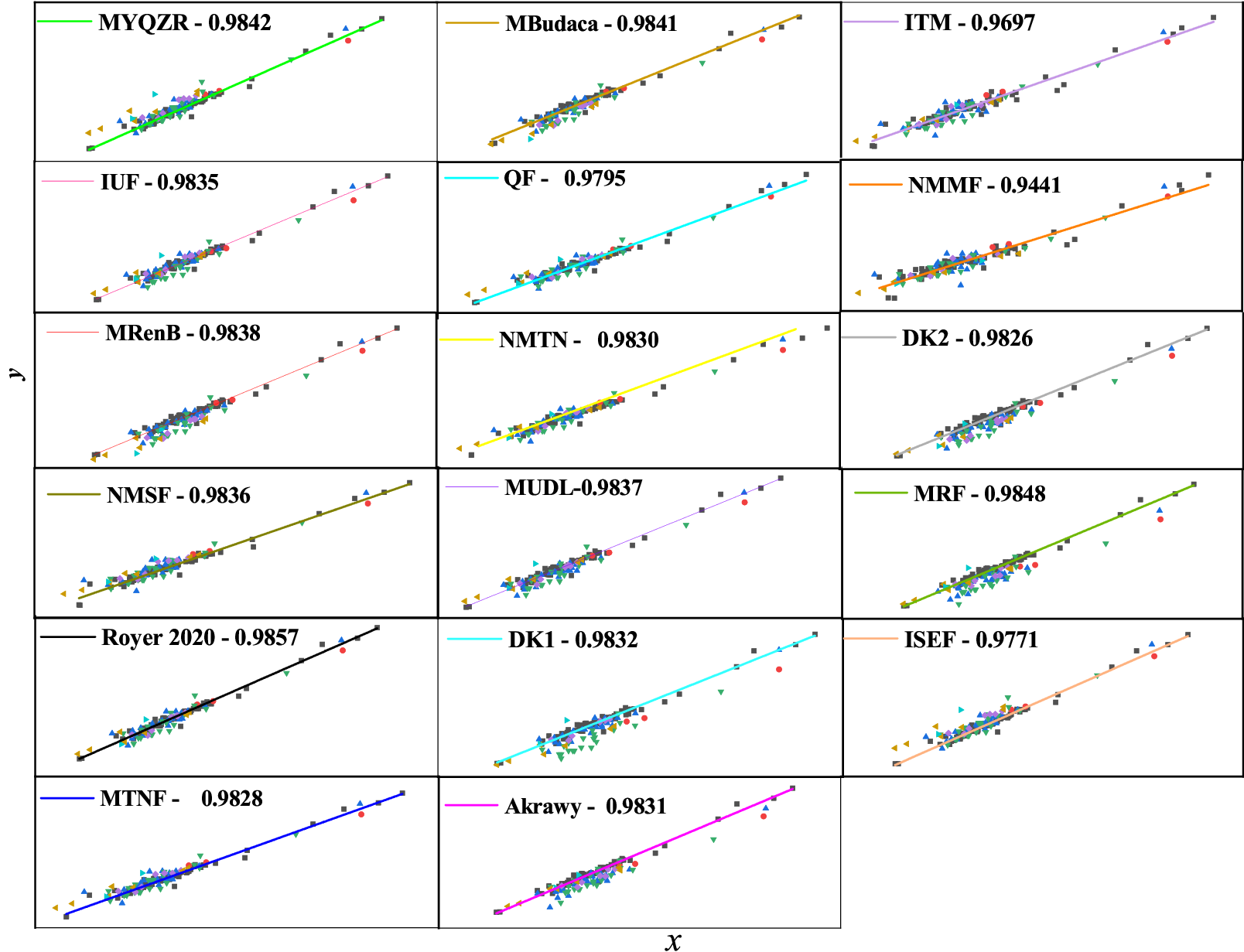}
\caption{Same as Fig. \ref{linear_fittings} but for the remaining 17 formulas from Table \ref{RMSE_table}. The corresponding $R^2$ value is indicated in each panel. Details of the $x$ and $y$ variables used, are provided in Table~\ref{coefficientISE} of Appendix~B.
}
\label{linear_fitting}
\end{figure*}
On the other hand, similar plots corresponding to the remaining 17 formulas are presented in Fig.~\ref{linear_fitting}, with the respective $R^2$ values indicated in each panel. Details of the $x$ and $y$ variables used are provided in Table~\ref{coefficientISE} of Appendix~B. The fitting parameters which we have used in this linear fitting for all 17 formulas are refitted using the current isomeric dataset.\par

It reflects the better linear dependence of experimental $\alpha$-decay half-lives on $Q_{\alpha}$ dependent terms for the NMHF formula with present fitting while removing the contribution of $l$ on the $\alpha$-decay half-lives. In other words, the NMHF formula with present fitting incorporates the dependence on $l$ more effectively. This $l$-dependence is critical for accurately predicting the half-lives of $\alpha$-decay processes involving isomeric states. Consequently, the NMHF formula with present fitting is a suitable tool for estimating the half-lives of $\alpha$-decay in isomers where experimental data is unavailable.\par

\begin{table*}[!htbp]
\caption{Evaluation of theoretical and experimental $\alpha$-decay half-lives for even-A nuclei either originating from or residing in an isomeric state. The experimental data of half-lives and spin parities are taken from NUBASE2020 \cite{audi20201} and NNDC \cite{nndc}. Q$_{\alpha}$ values are also experimental and taken from AME2020 \cite{audi20202}.}
\centering
\def\arraystretch{1.15}
\resizebox{1.00\textwidth}{!}{%
{\begin{tabular}{ccccccc|ccccccc}
 \hline
\multicolumn{1}{c}{$\alpha$-transition}&
\multicolumn{1}{c}{$Q_{\alpha}$}&
\multicolumn{1}{c}{j$_{p}$}&
\multicolumn{1}{c}{j$_{d}$}&
\multicolumn{1}{c}{$l$}&
\multicolumn{2}{c|}{$log_{10} T_{1/2}$(sec.)}&
\multicolumn{1}{c}{$\alpha$-transition}&
\multicolumn{1}{c}{$Q_{\alpha}$}&
\multicolumn{1}{c}{j$_{p}$}&
\multicolumn{1}{c}{j$_{d}$}&
\multicolumn{1}{c}{$l$}&
\multicolumn{2}{c}{$log_{10}T_{1/2}$(sec.)}\\
\cline{6-7}\cline{13-14}
\multicolumn{1}{c}{}&
\multicolumn{1}{c}{(MeV)}&
\multicolumn{1}{c}{}&
\multicolumn{1}{c}{}&
\multicolumn{1}{c}{}&
\multicolumn{1}{c}{Expt.}&
\multicolumn{1}{c|}{NMHF}&
\multicolumn{1}{c}{}&
\multicolumn{1}{c}{(MeV)}&
\multicolumn{1}{c}{}&
\multicolumn{1}{c}{}&
\multicolumn{1}{c}{}&
\multicolumn{1}{c}{Expt.}&
\multicolumn{1}{c}{NMHF}\\
 \hline
\multicolumn{14}{c}{Ground to Isomeric transition}\\
\hline
$^{154}$Ho$\rightarrow$$^{150}$Tb$^{m}$& 4.04 & 2   $^{-}$&2   $^{-}$ & 0 &6.57&5.95  & $^{194}$At$\rightarrow$$^{190}$Bi$^{m}$&7.33 & 5 $^{-}$&5  $^{-}$&0  &-0.54&-0.99  \\
$^{170}$Ir$\rightarrow$$^{166}$Re$^{p}$& 5.83 & 3   $^{-}$&3   $^{-}$ & 0 &1.22&1.34  & $^{220}$At$\rightarrow$$^{216}$Bi$^{m}$&6.05 & 3 $^{-}$&3  $^{-}$&0  & 3.44&4.34\\
$^{180}$Tl$\rightarrow$$^{176}$Au$^{m}$& 6.57 & 4   $^{-}$&7   $^{+}$ & 3 &1.19&0.56  & $^{240}$Am$\rightarrow$$^{236}$Np$^{p}$&5.47 & 3 $^{-}$&3  $^{-}$&0  &10.98&11.44 \\
$^{196}$Bi$\rightarrow$$^{192}$Tl$^{p}$& 5.26 & 3   $^{+}$&3   $^{+}$ & 0 &7.43&6.84  & $^{246}$Es$\rightarrow$$^{242}$Bk$^{p}$&7.49 & 4 $^{-}$&4  $^{-}$&0  & 3.66&3.62 \\
\hline
\multicolumn{14}{c}{Isomeric to Ground transition}\\
 \hline
$^{170}$Ir$^{m}$$\rightarrow$$^{166}$Re$^{ }$& 6.27 & 8$^{+}$& 7$^{+}$ & 2 & 0.33&-0.05  & $^{214}$Fr$^{m}$$\rightarrow$$^{210}$At$^{}$&8.71&8$^{-}$&5$^{+}$ &3 &-2.47&-3.62 \\
$^{172}$Ir$^{m}$$\rightarrow$$^{168}$Re$^{ }$& 6.13 & 7$^{+}$& 7$^{+}$ & 0 & 1.36&0.28  & $^{216}$Ac$^{m}$$\rightarrow$$^{212}$Fr$^{}$&9.28&9$^{-}$&5$^{+}$ &5 &-3.36 &-2.97 \\
$^{174}$Ir$^{m}$$\rightarrow$$^{170}$Re$^{ }$& 5.82 & 7$^{+}$& 5$^{+}$ & 2 & 2.30&1.79  & $^{242}$Am$^{m}$$\rightarrow$$^{238}$Np$^{}$&5.64&5$^{-}$&2$^{+}$ &3 &11.99 &11.04 \\
$^{186}$Tl$^{m}$$\rightarrow$$^{182}$Au$^{ }$& 6.02 & 7$^{+}$& 2$^{+}$ & 6 & 5.66&4.57  & $^{254}$Es$^{m}$$\rightarrow$$^{250}$Bk$^{}$&6.70&2$^{+}$&2$^{-}$ &1 & 7.65 &7.06 \\
$^{212}$Bi$^{m}$$\rightarrow$$^{208}$Tl$^{ }$& 6.46 & 9$^{-}$& 5$^{+}$ & 5 & 3.35&4.12  & $^{256}$Md$^{m}$$\rightarrow$$^{252}$Es$^{}$&7.90&7$^{-}$&4$^{+}$ &3 & 4.70 &3.40 \\
\hline
\multicolumn{14}{c}{Isomeric to Isomeric transition}\\
\hline
$^{152}$Ho$^{m}$$\rightarrow$$^{148}$Tb$^{m}$& 4.58 & 9$^{+}$& 9$^{+}$ &0 & 2.67&3.00  &$^{192}$Bi$^{m}$$\rightarrow$$^{188}$Tl$^{m}$&6.49&10$^{-}$& 7$^{+}$ &3 & 2.60&1.80\\
$^{154}$Ho$^{m}$$\rightarrow$$^{150}$Tb$^{m}$& 3.82 & 8$^{+}$& 9$^{+}$ &2 & 7.27&7.53  &$^{194}$Bi$^{n}$$\rightarrow$$^{190}$Tl$^{m}$&6.02&10$^{-}$& 7$^{+}$ &3 & 4.76&3.72\\
$^{154}$Tm$^{m}$$\rightarrow$$^{150}$Ho$^{m}$& 5.16 & 9$^{+}$& 9$^{+}$ &0 & 0.76&1.08  &$^{196}$Bi$^{n}$$\rightarrow$$^{192}$Tl$^{n}$&5.26&10$^{-}$& 8$^{-}$ &2 & 7.80&7.07\\
$^{156}$Lu$^{m}$$\rightarrow$$^{152}$Tm$^{m}$& 5.71 & 9$^{+}$& 9$^{+}$ &0 &-0.70&-0.40 & $^{192}$At$^{m}$$\rightarrow$$^{188}$Bi$^{m}$&7.63&9$^{-}$& 7$^{+}$ &3 &-1.06&-1.46 \\
$^{158}$Ta$^{m}$$\rightarrow$$^{154}$Lu$^{m}$& 6.21 & 9$^{+}$& 9$^{+}$ &0 &-1.43&-1.55 &$^{194}$At$^{m}$$\rightarrow$$^{190}$Bi$^{m}$&7.31&10$^{-}$&10$^{-}$ &0 &-0.49&-0.93 \\
$^{158}$Ta$^{n}$$\rightarrow$$^{154}$Lu$^{m}$& 8.87 &19$^{-}$& 9$^{+}$&11 &-3.36&-2.86 &$^{198}$At$^{m}$$\rightarrow$$^{194}$Bi$^{m}$&6.99&10$^{-}$&10$^{-}$ &0 & 0.13&0.23 \\
$^{160}$Ta$^{m}$$\rightarrow$$^{156}$Lu$^{m}$& 5.55 &10$^{+}$&10$^{+}$ &0 & 0.65&0.90  & $^{200}$At$^{m}$$\rightarrow$$^{196}$Bi$^{m}$&6.53&7$^{+}$& 7$^{+}$ &0 & 2.04&1.95 \\
$^{162}$Re$^{m}$$\rightarrow$$^{158}$Ta$^{m}$& 6.27 & 9$^{+}$& 9$^{+}$ &0 &-1.07&-1.12 &$^{200}$At$^{n}$$\rightarrow$$^{196}$Bi$^{m}$&6.76&10$^{-}$& 7$^{+}$ &3 & 1.52&1.58\\
$^{164}$Re$^{m}$$\rightarrow$$^{160}$Ta$^{m}$& 5.77 & 9$^{+}$& 9$^{+}$ &0 & 1.47&0.78  & $^{202}$At$^{m}$$\rightarrow$$^{198}$Bi$^{m}$&6.25&7$^{+}$& 7$^{+}$ &0 & 3.33&3.12\\
$^{164}$Ir$^{m}$$\rightarrow$$^{160}$Re$^{m}$& 7.06 & 9$^{+}$& 9$^{+}$ &0 &-2.76&-3.03 &$^{202}$At$^{n}$$\rightarrow$$^{198}$Bi$^{n}$&6.40&10$^{-}$&10$^{-}$ &0 & 2.68&2.51\\
$^{166}$Ir$^{m}$$\rightarrow$$^{162}$Re$^{m}$& 6.72 & 9$^{+}$& 9$^{+}$ &0 &-1.81&-1.94 &$^{200}$Fr$^{m}$$\rightarrow$$^{196}$At$^{m}$&7.59&10$^{-}$& 5$^{+}$ &5 &-0.72&0.12\\
$^{168}$Ir$^{m}$$\rightarrow$$^{164}$Re$^{m}$& 6.47 & 9$^{+}$& 9$^{+}$ &0 &-0.67&-1.06 & $^{204}$Fr$^{m}$$\rightarrow$$^{200}$At$^{m}$&7.11&7$^{+}$& 7$^{+}$ &0 & 0.41&0.56\\
$^{170}$Au$^{m}$$\rightarrow$$^{166}$Ir$^{m}$& 7.27 & 9$^{+}$& 9$^{+}$ &0 &-2.83&-2.99 &$^{204}$Fr$^{n}$$\rightarrow$$^{200}$At$^{n}$&7.15&10$^{-}$&10$^{-}$ &0 & 0.49&0.40\\
$^{176}$Au$^{m}$$\rightarrow$$^{172}$Ir$^{m}$& 6.43 & 8$^{+}$& 7$^{+}$ &2 & 0.13&0.12  & $^{206}$Fr$^{m}$$\rightarrow$$^{202}$At$^{m}$&6.93&7$^{+}$& 7$^{+}$ &0 & 1.28&1.22\\
$^{178}$Au$^{n}$$\rightarrow$$^{174}$Ir$^{m}$& 6.05 & 7$^{+}$& 7$^{+}$ &0 & 0.43&1.36  &$^{206}$Fr$^{m}$$\rightarrow$$^{202}$At$^{n}$&7.06&10$^{-}$&10$^{-}$ &0 & 0.73&0.75 \\
$^{186}$Bi$^{m}$$\rightarrow$$^{182}$Tl$^{m}$& 7.88 &10$^{-}$& 7$^{+}$ &3 &-2.01&-2.81 & $^{218}$Fr$^{m}$$\rightarrow$$^{214}$At$^{n}$&7.87&8$^{-}$& 9$^{-}$ &2 &-1.66&-1.46 \\
$^{188}$Bi$^{n}$$\rightarrow$$^{184}$Tl$^{m}$& 6.97 &10$^{-}$&10$^{-}$ &0 &-0.58&-0.50 &$^{206}$Ac$^{m}$$\rightarrow$$^{202}$Fr$^{m}$&7.90&10$^{-}$&10$^{-}$ &0 &-1.39&-1.40 \\
$^{190}$Bi$^{m}$$\rightarrow$$^{186}$Tl$^{m}$& 7.01 &10$^{-}$& 7$^{+}$ &3 & 0.95&-0.10 &                                              &    &       &         &  &     & \\
\hline
\end{tabular}}
}
\label{even-A}
\end{table*}

The experimental data for these isomeric $\alpha$-transitions are tabulated in Tables \ref{even-A}, \ref{odd-A}, and \ref{odd}. Columns 1 and 8 denote the transitions of $\alpha$-decay from or into any one of the isomeric states within the range 67$\leq$Z$\leq$110. The superscripts m, n, and p in the nuclide names denote the first, second, and third isomeric states of the nuclei (isomers), respectively.

Columns 2 and 9 display the disintegration energy of these types of isomeric $\alpha$-decays, calculated using Eqn. (\ref{Q}). Spin-parities, taken from NUBASE2020 \cite{audi20201}, NNDC \cite{nndc} or Ref. \cite{moller2019}, are listed in columns 3, 10, and 4, 11 for the parent nuclei and daughter nuclei, respectively. The $l$, calculated using the selection rule, is tabulated in columns 5 and 12. It's evident from this table that the results obtained using the NMHF with present fitting (columns 7 and 14) closely align with the experimental findings (as indicated in columns 6 and 13). It is to be pointed out that, higher value to $l$ contributes less to the decay rate \cite{Schuurmans2000,Severijns2005,Krause1998}, and are tough to be predicted. So only the lower $l$ values are of importance. In general, isomers in superheavy region are long-lived and can be observed when they decay to the any isomeric state with the lower $Q_{\alpha}$ value and some $l$ transfer as we can see in the Tables \ref{even-A}, \ref{odd-A}, and \ref{odd}. However, it is noteworthy that in most cases, the $\alpha$-decay half-lives ($T_{1/2}$) of transitions from isomeric states to the ground state are shorter than those from the corresponding ground states, primarily due to the relatively higher $Q_{\alpha}$ values of these isomers.
\begin{table*}[!htbp]
\caption{Same as Table \ref{even-A} but for Odd-A.}
\centering
\def\arraystretch{1.15}
\resizebox{1.00\textwidth}{!}{%
{\begin{tabular}{ccccccc|ccccccc}
 \hline
\multicolumn{1}{c}{$\alpha$-transition}&
\multicolumn{1}{c}{$Q_{\alpha}$}&
\multicolumn{1}{c}{j$_{p}$}&
\multicolumn{1}{c}{j$_{d}$}&
\multicolumn{1}{c}{$l$}&
\multicolumn{2}{c|}{$log_{10} T_{1/2}$(sec.)}&
\multicolumn{1}{c}{$\alpha$-transition}&
\multicolumn{1}{c}{$Q_{\alpha}$}&
\multicolumn{1}{c}{j$_{p}$}&
\multicolumn{1}{c}{j$_{d}$}&
\multicolumn{1}{c}{$l$}&
\multicolumn{2}{c}{$log_{10}T_{1/2}$(sec.)}\\
\cline{6-7}\cline{13-14}
\multicolumn{1}{c}{}&
\multicolumn{1}{c}{(MeV)}&
\multicolumn{1}{c}{}&
\multicolumn{1}{c}{}&
\multicolumn{1}{c}{}&
\multicolumn{1}{c}{Expt.}&
\multicolumn{1}{c|}{NMHF}&
\multicolumn{1}{c}{}&
\multicolumn{1}{c}{(MeV)}&
\multicolumn{1}{c}{}&
\multicolumn{1}{c}{}&
\multicolumn{1}{c}{}&
\multicolumn{1}{c}{Expt.}&
\multicolumn{1}{c}{NMHF}\\
 \hline
\multicolumn{14}{c}{Ground to Isomeric transition}\\
\hline
$^{151}$Ho$\rightarrow$$^{147}$Tb$^{m}$& 4.64 &11/2$^{-}$&11/2$^{-}$ & 0 & 2.20&2.63 & $^{193}$At$\rightarrow$$^{189}$Bi$^{m}$& 7.39& 1/2$^{+}$&1/2$^{+}$ &0 &-1.55&-1.19 \\
$^{153}$Ho$\rightarrow$$^{149}$Tb$^{m}$& 4.02 &11/2$^{-}$&11/2$^{-}$ & 0 & 5.37&6.04 & $^{195}$At$\rightarrow$$^{191}$Bi$^{m}$& 7.10& 1/2$^{+}$&1/2$^{+}$ &0 &-0.54&-0.21 \\
$^{157}$Lu$\rightarrow$$^{153}$Tm$^{m}$& 5.06 & 1/2$^{+}$& 1/2$^{+}$ & 0 & 3.83&2.22 & $^{197}$Fr$\rightarrow$$^{193}$At$^{m}$& 7.90& 7/2$^{-}$&7/2$^{-}$ &0 &-2.64&-2.10 \\
$^{157}$Ta$\rightarrow$$^{153}$Lu$^{m}$& 6.27 & 1/2$^{+}$& 1/2$^{+}$ & 0 &-1.98&-1.80& $^{205}$Ac$\rightarrow$$^{201}$Fr$^{m}$& 7.80&9/2$^{-}$&13/2$^{+}$ &3 &-1.10&-0.62\\
$^{159}$Ta$\rightarrow$$^{155}$Lu$^{m}$& 5.66 & 1/2$^{+}$& 1/2$^{+}$ & 0 & 0.49&0.43 & $^{241}$Cf$\rightarrow$$^{237}$Cm$^{m}$& 7.47& 7/2$^{-}$&7/2$^{-}$ &0 & 2.96&3.32\\
$^{171}$Ir$\rightarrow$$^{167}$Re$^{m}$& 5.87 & 1/2$^{+}$& 1/2$^{+}$ & 0 & 1.33&1.24 & $^{253}$Cf$\rightarrow$$^{249}$Cm$^{m}$& 6.08& 7/2$^{+}$&7/2$^{+}$ &0 & 8.70&9.57\\
$^{179}$Pb$\rightarrow$$^{175}$Hg$^{m}$& 7.10 & 9/2$^{-}$&13/2$^{+}$ & 3 &-2.57&-0.98& $^{243}$Es$\rightarrow$$^{239}$Bk$^{p}$& 8.03& 7/2$^{+}$&3/2$^{-}$ &3 & 1.58&2.16\\
$^{181}$Pb$\rightarrow$$^{177}$Hg$^{m}$& 6.92 & 9/2$^{-}$&13/2$^{+}$ & 3 &-1.41&-0.32& $^{245}$Es$\rightarrow$$^{241}$Bk$^{p}$& 7.86& 3/2$^{-}$&3/2$^{-}$ &0 & 2.13&2.29\\
$^{183}$Pb$\rightarrow$$^{179}$Hg$^{m}$& 6.76 & 3/2$^{-}$&13/2$^{+}$ & 5 &-0.27&1.02 & $^{247}$Es$\rightarrow$$^{243}$Bk$^{p}$& 7.44& 7/2$^{+}$&7/2$^{-}$ &1 & 3.59&3.89\\
$^{185}$Pb$\rightarrow$$^{181}$Hg$^{m}$& 6.63 & 3/2$^{-}$& 3/2$^{-}$ & 0 & 1.27&0.29 & $^{249}$Es$\rightarrow$$^{245}$Bk$^{p}$& 6.89& 7/2$^{+}$&7/2$^{+}$ &0 & 6.03&6.05\\
$^{187}$Pb$\rightarrow$$^{183}$Hg$^{m}$& 6.21 &13/2$^{+}$&13/2$^{+}$ & 0 & 2.18&1.92 & $^{255}$Es$\rightarrow$$^{251}$Bk$^{m}$& 6.40& 7/2$^{+}$&7/2$^{+}$ &0 & 7.63&8.36\\
$^{187}$Bi$\rightarrow$$^{183}$Tl$^{m}$& 7.15 & 9/2$^{-}$& 9/2$^{-}$ & 0 &-1.42&-1.12& $^{253}$Fm$\rightarrow$$^{249}$Cf$^{m}$& 7.05& 1/2$^{+}$&5/2$^{+}$ &2 & 6.33&6.02\\
$^{189}$Bi$\rightarrow$$^{185}$Tl$^{m}$& 6.81 & 9/2$^{-}$& 9/2$^{-}$ & 0 &-0.16&0.07 & $^{255}$No$\rightarrow$$^{251}$Fm$^{m}$& 8.23& 1/2$^{+}$&5/2$^{+}$ &2 & 2.85&2.32 \\
$^{191}$Bi$\rightarrow$$^{187}$Tl$^{m}$& 6.44 & 9/2$^{-}$& 9/2$^{-}$ & 0 & 1.39&1.44 & $^{255}$Lr$\rightarrow$$^{251}$Md$^{p}$& 8.50& 1/2$^{-}$&1/2$^{-}$ &0 & 1.49&1.47 \\
$^{193}$Bi$\rightarrow$$^{189}$Tl$^{m}$& 6.03 & 9/2$^{-}$& 9/2$^{-}$ & 0 & 3.46&3.16 & $^{259}$Lr$\rightarrow$$^{255}$Md$^{p}$& 8.57& 1/2$^{-}$&1/2$^{-}$ &0 & 0.90&1.30\\
$^{195}$Bi$\rightarrow$$^{191}$Tl$^{m}$& 5.54 & 9/2$^{-}$& 9/2$^{-}$ & 0 & 5.79&5.43 & $^{259}$Rf$\rightarrow$$^{255}$No$^{m}$& 9.03& 7/2$^{+}$&7/2$^{+}$ &0 & 0.42&0.19 \\
$^{197}$Bi$\rightarrow$$^{193}$Tl$^{m}$& 4.99 & 9/2$^{-}$& 9/2$^{-}$ & 0 & 8.75&8.31 & $^{259}$Db$\rightarrow$$^{255}$Lr$^{m}$& 9.58& 9/2$^{+}$&7/2$^{-}$ &1 &-0.29&-1.01  \\
$^{209}$Po$\rightarrow$$^{205}$Pb$^{m}$& 4.98 & 1/2$^{-}$& 1/2$^{-}$ & 0 & 9.59&9.20 & $^{269}$Ds$\rightarrow$$^{265}$Hs$^{m}$&11.28& 9/2$^{+}$&9/2$^{+}$ &0 &-3.64&-3.88  \\
$^{191}$At$\rightarrow$$^{187}$Bi$^{m}$& 7.71 & 1/2$^{+}$& 1/2$^{+}$ & 0 &-2.77&-2.21& $^{277}$Cn$\rightarrow$$^{273}$Ds$^{m}$&11.42& 3/2$^{+}$&3/2$^{+}$ &0 &-3.10&-3.65  \\

\hline
\multicolumn{14}{c}{Isomeric to Ground transition}\\
\hline
$^{151}$Ho$^{m}$$\rightarrow$$^{147}$Tb& 4.74 & 1/2$^{+}$& 1/2$^{+}$ & 0 & 1.79&2.20 & $^{155}$Lu$^{m}$$\rightarrow$$^{151}$Tm& 7.58&25/2$^{-}$&11/2 $^{-}$ & 8 &-2.57&-2.99  \\
$^{153}$Ho$^{m}$$\rightarrow$$^{149}$Tb& 4.12 & 1/2$^{+}$& 1/2$^{+}$ & 0 & 5.49&5.43 & $^{157}$Ta$^{m}$$\rightarrow$$^{153}$Lu& 7.94&25/2$^{-}$&11/2 $^{-}$ & 8 &-2.77&-3.37  \\
$^{157}$Lu$^{m}$$\rightarrow$$^{153}$Tm& 5.13 &11/2$^{-}$&11/2$^{-}$ & 0 & 1.90&1.94 & $^{159}$Re$^{m}$$\rightarrow$$^{155}$Ta& 6.97&11/2$^{-}$&11/2 $^{-}$ & 0 &-3.57&-3.40  \\
$^{159}$Ta$^{m}$$\rightarrow$$^{155}$Lu& 5.75 &11/2$^{-}$&11/2$^{-}$ & 0 &-0.03&0.11 & $^{167}$Re$^{m}$$\rightarrow$$^{163}$Ta& 5.41& 1/2 $^{+}$&1/2 $^{+}$ & 0 & 2.77&2.36 \\
$^{185}$Hg$^{m}$$\rightarrow$$^{181}$Pt& 5.87 &13/2$^{+}$& 1/2$^{-}$ & 7 & 4.86&4.91 & $^{173}$Ir$^{m}$$\rightarrow$$^{169}$Re& 5.94&11/2 $^{-}$&9/2 $^{-}$ & 2 & 1.29&1.27 \\
$^{183}$Pb$^{m}$$\rightarrow$$^{179}$Hg& 7.03 &13/2$^{+}$& 7/2$^{-}$ & 3 &-0.38&-0.61& $^{187}$Tl$^{m}$$\rightarrow$$^{183}$Au& 5.66& 9/2 $^{-}$&5/2 $^{-}$ & 2 & 4.02&4.18 \\
$^{185}$Bi$^{m}$$\rightarrow$$^{181}$Tl& 8.22 & 1/2$^{+}$& 1/2$^{+}$ & 0 &-3.24&-4.24& $^{191}$Bi$^{m}$$\rightarrow$$^{187}$Tl& 7.02& 1/2 $^{+}$&1/2 $^{+}$ & 0 &-0.74&-0.59  \\
$^{187}$Bi$^{m}$$\rightarrow$$^{183}$Tl& 7.89 & 1/2$^{+}$& 1/2$^{+}$ & 0 &-3.43&-3.32& $^{193}$Bi$^{m}$$\rightarrow$$^{189}$Tl& 6.62& 1/2 $^{+}$&1/2 $^{+}$ & 0 & 0.58&0.86 \\
$^{191}$At$^{m}$$\rightarrow$$^{187}$Bi& 7.88 & 7/2$^{-}$& 9/2$^{-}$ & 2 &-2.68&-2.44& $^{195}$Bi$^{m}$$\rightarrow$$^{191}$Tl& 6.23& 1/2 $^{+}$&1/2 $^{+}$ & 0 & 2.42&2.38 \\
$^{193}$At$^{m}$$\rightarrow$$^{189}$Bi& 7.58 & 7/2$^{-}$& 9/2$^{-}$ & 2 &-1.68&-1.52& $^{203}$Po$^{m}$$\rightarrow$$^{199}$Pb& 6.14&13/2 $^{+}$&3/2 $^{-}$ & 5 & 5.05&4.52 \\
$^{193}$At$^{n}$$\rightarrow$$^{189}$Bi& 7.61 &13/2$^{+}$& 9/2$^{-}$ & 3 &-0.95&-1.38& $^{211}$Po$^{m}$$\rightarrow$$^{207}$Pb& 9.06&25/2 $^{+}$&1/2 $^{-}$ &13 & 1.40&1.63 \\
$^{247}$Fm$^{m}$$\rightarrow$$^{243}$Cf& 8.30 & 1/2$^{+}$& 1/2$^{+}$ & 0 & 0.77&1.13 & $^{199}$At$^{m}$$\rightarrow$$^{195}$Bi& 7.02& 1/2 $^{+}$&9/2 $^{-}$ & 5 & 1.49&1.40 \\
$^{249}$Md$^{m}$$\rightarrow$$^{245}$Es& 8.54 & 1/2$^{-}$& 3/2$^{-}$ & 2 & 0.28&0.94 & $^{217}$Ac$^{m}$$\rightarrow$$^{213}$Fr&11.84&29/2 $^{+}$&9/2 $^{-}$ &11 &-4.78&-4.71  \\
$^{257}$Rf$^{m}$$\rightarrow$$^{253}$No& 9.16 &11/2$^{-}$& 9/2$^{-}$ & 2 & 0.67&0.04 & $^{247}$Md$^{m}$$\rightarrow$$^{243}$Es& 9.02& 1/2 $^{-}$&7/2 $^{+}$ & 3 &-0.50&-0.28 \\
$^{271}$Ds$^{m}$$\rightarrow$$^{267}$Hs& 10.94& 9/2$^{+}$& 5/2$^{+}$ & 2 &-2.77&-2.84&                                        &     &           &          &  &     &       \\
\hline
\multicolumn{14}{c}{Isomeric to Isomeric transition}\\
\hline
$^{153}$Tm$^{m}$$\rightarrow$$^{149}$Ho$^{m}$& 5.24 & 1/2 $^{+}$& 1/2$^{+}$ & 0 & 0.43&0.71 & $^{187}$Pb$^{m}$$\rightarrow$$^{183}$Hg$^{m}$&6.24&3/2$^{-}$&13/2$^{+}$ &5& 2.34&3.02 \\
$^{155}$Tm$^{m}$$\rightarrow$$^{151}$Ho$^{m}$& 4.57 & 1/2 $^{+}$& 1/2$^{+}$ & 0 & 3.35&3.81 &$^{191}$Pb$^{m}$$\rightarrow$$^{187}$Hg$^{m}$&5.40&13/2$^{+}$&13/2$^{+}$ &0& 5.82&5.59 \\
$^{155}$Lu$^{m}$$\rightarrow$$^{151}$Tm$^{m}$& 5.73 & 1/2 $^{+}$& 1/2$^{+}$ & 0 &-0.74&-0.52&$^{195}$Po$^{m}$$\rightarrow$$^{191}$Pb$^{m}$&6.84&13/2$^{+}$&13/2$^{+}$ &0& 0.33&0.40 \\
$^{161}$Ta$^{m}$$\rightarrow$$^{157}$Lu$^{m}$& 5.28 &11/2 $^{-}$&11/2$^{-}$ & 0 & 1.64&2.07 &$^{197}$Po$^{m}$$\rightarrow$$^{193}$Pb$^{m}$&6.55&13/2$^{+}$&13/2$^{+}$ &0& 1.49&1.51\\
$^{161}$Re$^{m}$$\rightarrow$$^{157}$Ta$^{m}$& 6.43 &11/2 $^{-}$&11/2$^{-}$ & 0 &-1.81&-1.68&$^{199}$Po$^{m}$$\rightarrow$$^{195}$Pb$^{m}$&6.18&13/2$^{+}$&13/2$^{+}$ &0& 2.81&3.00\\
$^{163}$Re$^{m}$$\rightarrow$$^{159}$Ta$^{m}$& 6.07 &11/2 $^{-}$&11/2$^{-}$ & 0 &-0.49&-0.38&$^{201}$Po$^{m}$$\rightarrow$$^{197}$Pb$^{m}$&5.90&13/2$^{+}$&13/2$^{+}$ &0& 4.27&4.25\\
$^{165}$Re$^{m}$$\rightarrow$$^{161}$Ta$^{m}$& 5.64 &11/2 $^{-}$&11/2$^{-}$ & 0 & 1.25&1.30 & $^{197}$At$^{m}$$\rightarrow$$^{193}$Bi$^{m}$&6.85&1/2$^{+}$& 1/2$^{+}$ &0& 0.30&0.71\\
$^{169}$Re$^{m}$$\rightarrow$$^{165}$Ta$^{m}$& 5.18 & 1/2 $^{+}$& 9/2$^{-}$ & 5 & 3.88&4.71 &$^{195}$Rn$^{m}$$\rightarrow$$^{191}$Po$^{m}$&7.71&13/2$^{+}$&13/2$^{+}$ &0&-2.30&-1.86  \\
$^{165}$Ir$^{m}$$\rightarrow$$^{161}$Re$^{m}$& 6.88 &11/2 $^{-}$&11/2$^{-}$ & 0 &-2.64&-2.48&$^{197}$Rn$^{m}$$\rightarrow$$^{193}$Po$^{m}$&7.51&13/2$^{+}$&13/2$^{+}$ &0&-1.62&-1.20  \\
$^{167}$Ir$^{m}$$\rightarrow$$^{163}$Re$^{m}$& 6.56 &11/2 $^{-}$&11/2$^{-}$ & 0 &-1.50&-1.41&$^{203}$Rn$^{m}$$\rightarrow$$^{199}$Po$^{m}$&6.68&13/2$^{+}$&13/2$^{+}$ &0&-1.49&1.77 \\
$^{169}$Ir$^{m}$$\rightarrow$$^{165}$Re$^{m}$& 6.25 &11/2 $^{-}$&11/2$^{-}$ & 0 &-0.45&-0.25& $^{199}$Fr$^{m}$$\rightarrow$$^{195}$At$^{m}$&7.83&7/2$^{-}$& 7/2$^{-}$ &0&-2.19&-1.85 \\
$^{171}$Ir$^{m}$$\rightarrow$$^{167}$Re$^{m}$& 6.07 &11/2 $^{-}$&11/2$^{-}$ & 0 & 0.31&0.47 & $^{201}$Fr$^{m}$$\rightarrow$$^{197}$At$^{m}$&7.60&1/2$^{+}$& 1/2$^{+}$ &0&-1.62&-1.12  \\
$^{171}$Au$^{m}$$\rightarrow$$^{167}$Ir$^{m}$& 7.17 &11/2 $^{-}$&11/2$^{-}$ & 0 &-2.76&-2.66& $^{203}$Fr$^{m}$$\rightarrow$$^{199}$At$^{m}$&7.38&1/2$^{+}$& 1/2$^{+}$ &0&-0.67&-0.38  \\
$^{173}$Au$^{m}$$\rightarrow$$^{169}$Ir$^{m}$& 6.90 &11/2 $^{-}$&11/2$^{-}$ & 0 &-1.82&-1.77&$^{201}$Ra$^{m}$$\rightarrow$$^{197}$Rn$^{m}$&8.07&13/2$^{+}$&13/2$^{+}$ &0&-2.22&-2.25  \\
$^{175}$Au$^{m}$$\rightarrow$$^{171}$Ir$^{m}$& 6.58 &11/2 $^{-}$&11/2$^{-}$ & 0 & 2.18&-0.68&$^{203}$Ra$^{m}$$\rightarrow$$^{199}$Rn$^{m}$&7.78&13/2$^{+}$&13/2$^{+}$ &0&-1.62&-1.35 \\
$^{177}$Au$^{m}$$\rightarrow$$^{173}$Ir$^{m}$& 6.26 &11/2 $^{-}$&11/2$^{-}$ & 0 & 0.18&0.53 &$^{213}$Ra$^{m}$$\rightarrow$$^{209}$Rn$^{m}$&7.46&17/2$^{-}$&13/2$^{+}$ &3&-0.44&0.33\\

\hline
\end{tabular}}
}
\label{odd-A}
\end{table*}

\begin{table*}[!htbp]
\caption{Continuation of Table \ref{odd-A}.}
\centering
\def\arraystretch{1.15}
\resizebox{1.00\textwidth}{!}{%
{\begin{tabular}{ccccccc|ccccccc}
 \hline
\multicolumn{1}{c}{$\alpha$-transition}&
\multicolumn{1}{c}{$Q_{\alpha}$}&
\multicolumn{1}{c}{j$_{p}$}&
\multicolumn{1}{c}{j$_{d}$}&
\multicolumn{1}{c}{$l$}&
\multicolumn{2}{c|}{$log_{10} T_{1/2}$(sec.)}&
\multicolumn{1}{c}{$\alpha$-transition}&
\multicolumn{1}{c}{$Q_{\alpha}$}&
\multicolumn{1}{c}{j$_{p}$}&
\multicolumn{1}{c}{j$_{d}$}&
\multicolumn{1}{c}{$l$}&
\multicolumn{2}{c}{$log_{10}T_{1/2}$(sec.)}\\
\cline{6-7}\cline{13-14}
\multicolumn{1}{c}{}&
\multicolumn{1}{c}{(MeV)}&
\multicolumn{1}{c}{}&
\multicolumn{1}{c}{}&
\multicolumn{1}{c}{}&
\multicolumn{1}{c}{Expt.}&
\multicolumn{1}{c|}{NMHF}&
\multicolumn{1}{c}{}&
\multicolumn{1}{c}{(MeV)}&
\multicolumn{1}{c}{}&
\multicolumn{1}{c}{}&
\multicolumn{1}{c}{}&
\multicolumn{1}{c}{Expt.}&
\multicolumn{1}{c}{NMHF}\\
 \hline
\multicolumn{14}{c}{Isomeric to Isomeric transition}\\
\hline
$^{177}$Tl$^{m}$$\rightarrow$$^{173}$Au$^{m}$& 7.66 &11/2 $^{-}$&11/2$^{-}$ & 0 &-3.33&-3.41&$^{209}$Th$^{m}$$\rightarrow$$^{205}$Ra$^{m}$&8.34&13/2$^{+}$&13/2$^{+}$ &0&-2.60&-2.36 \\
$^{179}$Tl$^{m}$$\rightarrow$$^{175}$Au$^{m}$& 7.37 &11/2 $^{-}$&11/2$^{-}$ & 0 &-2.82&-2.53& $^{251}$No$^{m}$$\rightarrow$$^{247}$Fm$^{m}$&8.81&1/2$^{+}$& 1/2$^{+}$ &0& 0.01&0.18 \\
$^{179}$Tl$^{m}$$\rightarrow$$^{175}$Au$^{m}$& 7.38 &11/2 $^{-}$&11/2$^{-}$ & 0 &-2.82&-2.55& $^{253}$Lr$^{m}$$\rightarrow$$^{249}$Md$^{m}$&8.85&1/2$^{-}$& 1/2$^{-}$ &0& 0.17&0.38  \\
$^{181}$Tl$^{m}$$\rightarrow$$^{177}$Au$^{n}$& 6.73 & 9/2 $^{-}$& 9/2$^{-}$ & 0 &-0.46&-0.43& $^{257}$Db$^{m}$$\rightarrow$$^{253}$Lr$^{m}$&9.32&1/2$^{-}$& 1/2$^{-}$ &0&-0.11&-0.38  \\
$^{183}$Tl$^{m}$$\rightarrow$$^{179}$Au$^{p}$& 6.47 & 9/2 $^{-}$& 9/2$^{-}$ & 0 & 0.55&0.53 & $^{259}$Sg$^{m}$$\rightarrow$$^{255}$Rf$^{m}$&9.70&1/2$^{+}$& 5/2$^{+}$ &2&-0.63&-0.91  \\
$^{185}$Tl$^{m}$$\rightarrow$$^{181}$Au$^{m}$& 5.93 & 9/2 $^{-}$&11/2$^{-}$ & 2 & 2.06&2.91 & $^{263}$Sg$^{m}$$\rightarrow$$^{259}$Rf$^{m}$&9.24&7/2$^{+}$& 9/2$^{+}$ &2&-0.38&0.42 \\
$^{185}$Pb$^{m}$$\rightarrow$$^{181}$Hg$^{m}$& 6.56 &13/2 $^{+}$&13/2$^{+}$ & 0 & 0.93&0.56 &                                              &    &         &           & &     & \\
\hline
\end{tabular}}
}
\label{odd}
\end{table*}

For some cases, such as the transitions $^{186}$Tl$^{m} \rightarrow ^{182}$Au, $^{212}$Bi$^{m} \rightarrow ^{208}$Tl, $^{216}$Ac$^{m} \rightarrow ^{212}$Fr, $^{211}$Po$^{m} \rightarrow ^{207}$Pb, $^{217}$Ac$^{m} \rightarrow ^{213}$Fr, etc., the isomeric transitions have higher $Q_{\alpha}$ values and longer $T_{1/2}$ compared to the transitions from their ground states. This indicates that $\alpha$-decay half-lives, which primarily depend on $Q_{\alpha}$ values, are also significantly influenced by the $l$, which are 6, 5, 5, 13, and 11, respectively, in these particular cases. It is gratifying to note that the NMHF reproduces the experimental logarithmic half-lives of $\alpha$-decay in the isomers for all the $l$ values which establishes its applicability in isomeric state as well. \par
\begin{figure*}[!htbp]
\centering
\includegraphics[width=0.92\textwidth]{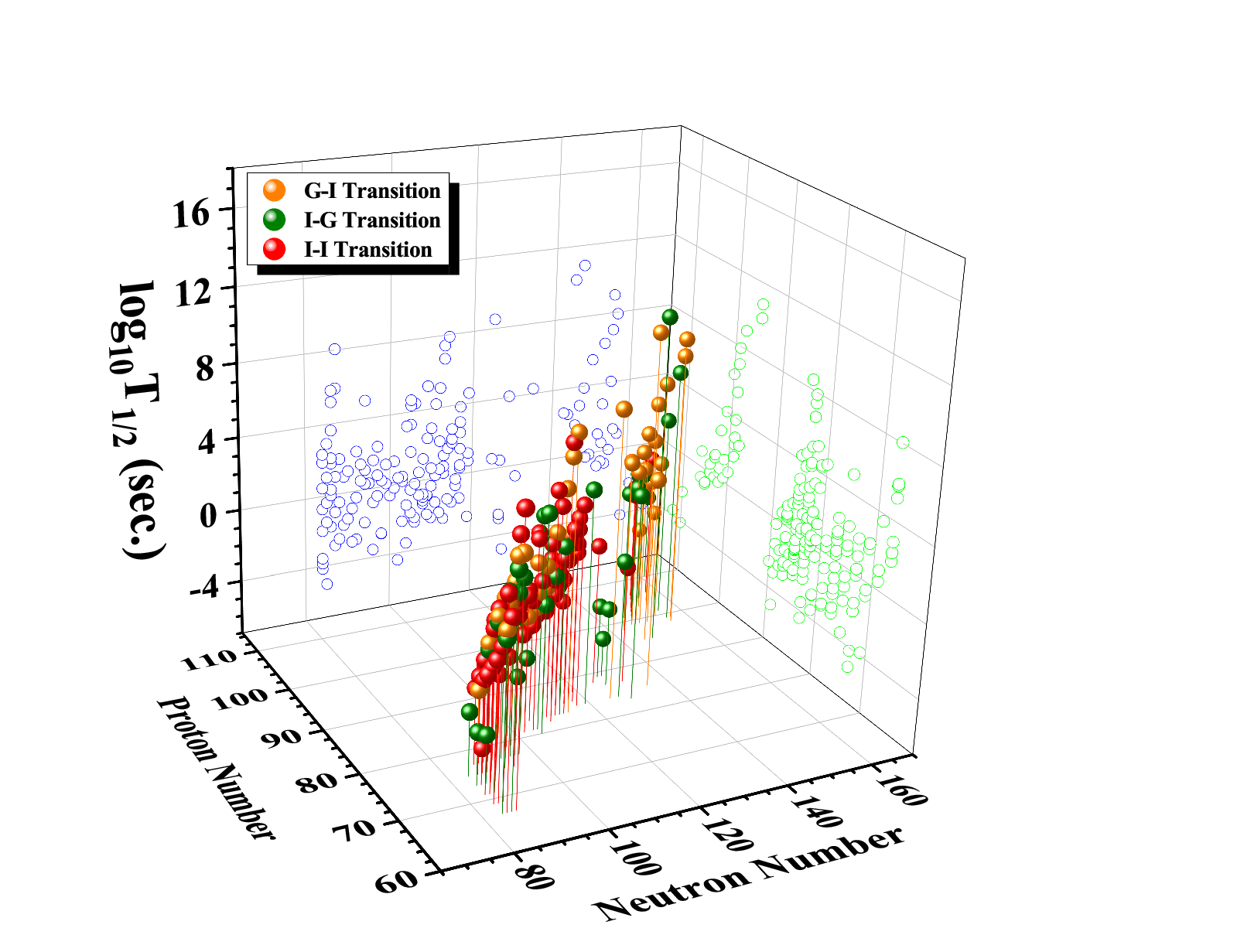}
\caption{(Colour online) Variation of logarithmic half-lives of $\alpha$-decay, calculated by using NMHF with present fitting for 165 experimental isomeric $\alpha$-transitions of ground to isomeric transition (G-I transition), isomeric to ground state transition (I-G transition) and isomeric to isomeric state transition (I-I transition).}
\label{three_dim.}
\end{figure*}

\begin{table*}[!htbp]
\caption{Predicted $\alpha$-decay half-lives for cases involving at least one decay from or in an isomeric state, which are
mentioned in NUBASE2020 \cite{audi20201}, using NMHF.}
\centering
\def\arraystretch{1.15}
\resizebox{1.00\textwidth}{!}{%
{\begin{tabular}{c@{\hskip 0.2in}c@{\hskip 0.2in}c@{\hskip 0.2in}c@{\hskip 0.2in}c@{\hskip 0.2in}c|c@{\hskip 0.2in}c@{\hskip
0.2in}c@{\hskip 0.2in}c@{\hskip 0.2in}c@{\hskip 0.3in}c}
 \hline
\multicolumn{1}{c}{$\alpha$}&
\multicolumn{1}{c}{$Q_{\alpha}$}&
 \multicolumn{1}{c}{j$_{p}$}&
\multicolumn{1}{c}{j$_{d}$}&
\multicolumn{1}{c}{$l_{min}$}&
 \multicolumn{1}{c|}{$log_{10}T_{\alpha}$}&
\multicolumn{1}{c}{$\alpha$}&
\multicolumn{1}{c}{$Q_{\alpha}$}&
 \multicolumn{1}{c}{j$_{p}$}&
\multicolumn{1}{c}{j$_{d}$}&
\multicolumn{1}{c}{$l_{min}$}&
 \multicolumn{1}{c}{$log_{10}T_{\alpha}$}\\
\multicolumn{1}{c}{Transition}&
\multicolumn{1}{c}{(MeV)}& \multicolumn{1}{c}{}& \multicolumn{1}{c}{}& \multicolumn{1}{c}{}& \multicolumn{1}{c|}{(sec.)}&\multicolumn{1}{c}{Transition}&
\multicolumn{1}{c}{(MeV)}& \multicolumn{1}{c}{}& \multicolumn{1}{c}{}& \multicolumn{1}{c}{}& \multicolumn{1}{c}{(sec.)}\\
 \hline
$^{179}$Au$^{p}$$\rightarrow$$^{175}$Ir$^{m}$&6.06 & 9/2$^{-}$& 9/2$^{-}$&0 &1.36 &$^{248}$Bk$^{p}$$\rightarrow$$^{244}$Am$^{ }$& 5.88&  5$^{-}$&  6$^{-}$&2&10.36\\
$^{181}$Au$^{p}$$\rightarrow$$^{177}$Ir$^{m}$&5.86 &11/2$^{-}$& 9/2$^{-}$&2 &2.47 &$^{250}$Bk$^{p}$$\rightarrow$$^{246}$Am$^{ }$& 5.74&  2$^{+}$&  7$^{-}$&5&12.12\\
$^{182}$Tl$^{p}$$\rightarrow$$^{178}$Au$^{n}$&6.87 &  10$^{-}$&   7$^{+}$&3 &-0.39&$^{241}$Cf$^{p}$$\rightarrow$$^{237}$Cm$^{ }$& 7.81&1/2$^{+}$&5/2$^{+}$&2&2.34\\
$^{192}$Tl$^{p}$$\rightarrow$$^{188}$Au$^{ }$&4.25 &   3$^{+}$&   1$^{-}$&3 &12.46&$^{245}$Es$^{p}$$\rightarrow$$^{241}$Bk$^{ }$& 8.20&7/2$^{-}$&7/2$^{+}$&1&1.22\\
$^{195}$At$^{p}$$\rightarrow$$^{191}$Bi$^{m}$&7.20 &13/2$^{+}$& 1/2$^{+}$&6 &1.18 &$^{245}$Es$^{q}$$\rightarrow$$^{241}$Bk$^{ }$& 8.25&1/2$^{-}$&7/2$^{+}$&3&1.47\\
$^{229}$Np$^{p}$$\rightarrow$$^{225}$Pa$^{ }$&7.18 & 5/2$^{-}$& 5/2$^{-}$&0 &2.61 &$^{246}$Es$^{p}$$\rightarrow$$^{242}$Bk$^{ }$& 7.84&  2$^{-}$&  3$^{+}$&1&2.44\\
$^{233}$Np$^{p}$$\rightarrow$$^{229}$Pa$^{ }$&5.68 & 5/2$^{-}$& 5/2$^{+}$&1 &9.44 &$^{255}$Fm$^{p}$$\rightarrow$$^{251}$Cf$^{ }$& 7.47&9/2$^{+}$&1/2$^{+}$&4&4.96\\
$^{236}$Np$^{p}$$\rightarrow$$^{232}$Pa$^{ }$&5.25 &   3$^{-}$&   2$^{-}$&2 &12.11&$^{257}$No$^{p}$$\rightarrow$$^{253}$Fm$^{m}$& 8.64&9/2$^{+}$&7/2$^{+}$&2&1.01\\
$^{243}$Np$^{p}$$\rightarrow$$^{239}$Pa$^{ }$&4.16 & 5/2$^{-}$& 1/2$^{+}$&3 &20.41&$^{259}$Rf$^{p}$$\rightarrow$$^{255}$No$^{p}$& 9.09&7/2$^{+}$&7/2$^{+}$&0&0.01\\
$^{237}$Cm$^{p}$$\rightarrow$$^{233}$Pu$^{ }$&6.98 & 7/2$^{-}$& 5/2$^{+}$&1 &4.56 &$^{259}$Rf$^{q}$$\rightarrow$$^{255}$No$^{m}$&9.13&9/2$^{+}$&11/2$^{-}$&1&-0.02\\
$^{243}$Cm$^{p}$$\rightarrow$$^{239}$Pu$^{ }$&6.27 & 7/2$^{+}$& 1/2$^{+}$&4 &8.51 &$^{262}$Sg$^{p}$$\rightarrow$$^{258}$Rf$^{ }$&10.46&  9$^{-}$& 0$^{+}$&10&1.46\\
$^{242}$Bk$^{p}$$\rightarrow$$^{238}$Am$^{ }$&7.06 &   4$^{-}$&   1$^{+}$&3 &5.07 & & & & &&\\

\hline
\end{tabular}}
}
\label{predicted}
\end{table*}

We have illustrated our findings in Fig. \ref{three_dim.}, depicting the calculated logarithmic half-lives of various $\alpha$-transitions by NMHF. These include 46 transitions from ground states to any isomeric state, 39 transitions from any isomeric state to ground states, and 80 transitions between isomeric states, presented in a three-dimensional plot. Our analysis reveals that the majority of isomers are concentrated in the heavy region, exhibiting half-lives ranging from 10$^{-5}$ to 10$^{+6}$ sec. Furthermore, a few isomers are situated in the superheavy region, characterized by relatively higher logarithmic half-lives, approximately around 10 sec. Notably, there is a conspicuous dearth of parent nuclei (particularly isomers) in the vicinity of N = 126.\par

\begin{figure*}[!htbp]
\centering
\includegraphics[width=0.7\textwidth]{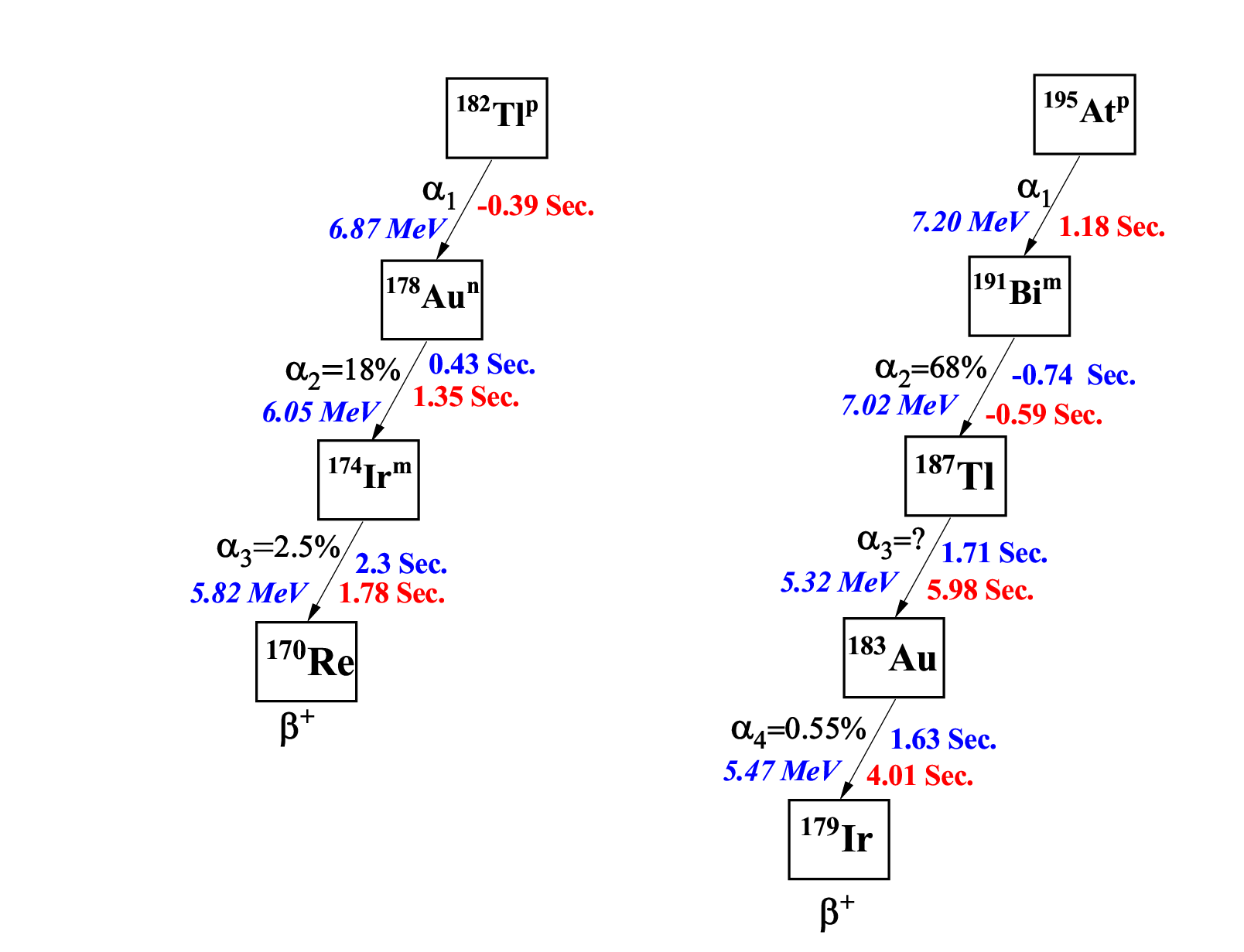}
\caption{The figure depicts the most probable $\alpha$-decay chains of heavy nuclei (isomers) $^{182}$Tl$^{p}$
and $^{195}$At$^{p}$, wherein the blue numbers represent experimental $Q_{\alpha}$-value and half-lives \cite{audi20201,nndc}, while the red numbers indicate theoretical half-lives calculated using the NMHF with present fitting. Experimental decay modes are shown in blue brackets.}
\label{decay_chain1}
\end{figure*}

We have employed NMHF in the unknown heavy and superheavy region to predict the half-lives for $\alpha$-transitions from isomeric state decay ranging from 79$\leq$Z$\leq$106. The selection criteria of these heavy and superheavy nuclei rely on their possible $\alpha$-transitions according to NUBASE2020 \cite{audi20201} which results a total of 23 nuclei under this category. For the estimation of half-lives of these 23 isomers, the spin-parities of parent and daughter nuclei are taken from NUBASE2020 \cite{audi20201}. Q$_{\alpha}$ are taken from AME2020 \cite{audi20202}. These predicted half-lives are listed in column 6 and 12 of Table \ref{predicted} which are found within the experimental reach and quite similar to the $\alpha$-decay half-lives of known isomers (see Tables \ref{even-A}, \ref{odd-A}, and \ref{odd}). Most of the nuclei in Table \ref{predicted} are the ones whose decay modes are yet unknown \cite{audi20201}. While isomers typically decay via internal conversion (IT), we have identified some finite chances of $\alpha$-decay in these isomers and calculated their $\alpha$-decay half-lives. For instance,  for $^{192}$Tl$^p$, with a 100 \% experimental likelihood of $\alpha$-decay without any measurement of half-life as mentioned in NUBASE2020 \cite{audi20201}, the predicted half-life is found to fall within the experimental range, providing valuable direction for future experiments. On the other hand, in some cases, such as $^{179}$Au$^p$ and $^{181}$Au$^p$, although IT is the dominant decay mode, but the possibilities of $\alpha$-decay remains nonzero, as indicated by the Q-values and predicted half-lives.

Finally, our investigation has identified probable $\alpha$-decay chains throughout the periodic table. We have presented two highly probable decay chains as representative examples from the heavy region, as illustrated in Fig. \ref{decay_chain1}. Among these decay chains, the initial nuclei $^{182}$Tl$^{p}$ and $^{195}$At$^{p}$ in the heavy region, are known according to NUBASE2020 \cite{audi20201} and NNDC \cite{nndc}, albeit with unavailable half-life data and decay modes. The selection of the energy levels (isomeric states) of the parent and daughter nuclei in these decay chains was carried out after a careful examination of the level schemes, absolute intensities, and excitation energies \cite{audi20201,nndc}. These factors provide essential insights into the transition probabilities of $\alpha$-decays, as well as other competing decay modes.



These chains consist of few isomeric $\alpha$-transitions where the half-life of the parent nucleus is unknown. In the heavy region, the first decay chain comprises $^{182}$Tl$^{p}$$\rightarrow$$^{178}$Au$^{n}$$\rightarrow$$^{174}$Ir$^{m}$$\rightarrow$$^{170}$Re, while the second decay chain involves $^{195}$At$^{p}$$\rightarrow$$^{191}$Bi$^{m}$$\rightarrow$$^{187}$Tl$\rightarrow$$^{183}$Au$\rightarrow$$^{179}$Ir. The theoretical half-life (represented in red) closely aligns with the available experimental half-lives (shown in blue). The Q$_\alpha$ values (indicated in blue) are also sourced from NUBASE2020 \cite{audi20201} and NNDC \cite{nndc}. Additionally, the probable decay modes along with their percentage are indicated in the parentheses to provide a better insight into the chances of $\alpha$-decay in the decay chain. This figure clearly shows that $^{191}$Bi$^{m}$ exhibits a highly probable $\alpha$-decay with a 65 \% intensity, while other nuclei primarily decay via $\beta^{+}$ emission. However, there are always some probabilities of undergoing $\alpha$-decay for which we have successfully calculated the half-lives, including those with unknown decay modes.

This study sheds light on new isomers within the heavy region, providing valuable insights for future experimental endeavors.\par


\section{Conclusion}
Utilizing the empirical NMHF formula, which is recalibrated using 165 data of isomeric $\alpha$-transitions, the calculation of $\alpha$-decay half-lives for isomeric states across heavy and superheavy nuclei with Z ranging from 67 to 110 has been conducted. The findings reveal a close agreement between the estimated $\alpha$-decay logarithmic half-lives and experimental data. Of particular note is the significant role played by the angular momentum-dependent term in determining the half-lives of $\alpha$-decay in isomeric states. Predictions have been made regarding the logarithmic half-lives of $\alpha$-transitions from various unknown heavy and superheavy isomers, leading to the identification of a total of 23 potential $\alpha$-decay isomers in the heavy and superheavy region according to NUBASE2020 \cite{audi20201} and NNDC \cite{nndc}. Additionally, two probable decay chains have been identified from the heavy region ($^{182}$Tl$^{p}$ and $^{195}$At$^{p}$).

\section{Acknowledgments}
 GS acknowledges the support provided by Department of Science \& Technology (DST), Government of Rajasthan, Jaipur, India under F24(1)DST/R \& D/2024-EAC-00378-6549873/819, and would like to thank Bhoomika Maheshwari, Grand Accélérateur National d'Ions Lourds, France. AJ is indebted to S.K. Jain, Manipal University Jaipur, Jaipur, India for the valuable communication.
\section{Appendix}
\subsection{Appendix-A}
The other 17 formulas and coefficients of these refitted formulas on the experimental data set belonging to isomeric $\alpha$-transitions:
\begin{enumerate}
\begin{small}
\item \textbf{MYQZR}\\
In 2019, Akrawy \textit{et al.} modified Yibin \textit{et al.} formula (MYQZR) by adding the two asymmetry dependent (I and I$^{2}$) named as MYQZR \cite{MYQZR2019}:
\begin{equation}
\log_{10} T_{1/2} = a \sqrt{\mu 2 Z_d} Q^{-1/2} + b \sqrt{\mu (2 Z_d)}
 + c I + d I^2+ e \frac{l(l+1)}{\mu \sqrt{2 Z_d} A_1^{1/6}} +f.
\end{equation}
The coefficients of the MYQZR formula are as follows:
for even-even nuclei: $a = 0.3901$, $b = -1.5159$, $c = 0.5940$, $d = 13.9854$, $e = 4.8900$, $f = -9.9663$;
for even-odd nuclei: $a = 0.3975$, $b = -1.4742$, $c = 40.0961$, $d = -106.7169$, $e = 4.9378$, $f = -15.3001$;
for odd-even nuclei: $a = 0.3908$, $b = -1.4522$, $c = -2.4343$, $d = 8.4916$, $e = 5.3311$, $f = -11.2961$;
and for odd-odd nuclei: $a = 0.4021$, $b = -1.3763$, $c = 5.8230$, $d = -17.9362$, $e = 5.0431$, $f = -14.7770$.
\item \textbf{IUF}\\
The Improved Unified Formula (IUF) is an enhanced iteration of the NRDX empirical formula for $\alpha$-decay half-lives, as proposed by Ismail \textit{et al.} in 2022 \cite{ismail2022improved}. This modification incorporates the influences of angular momentum, isospin asymmetry, and parity.
\begin{eqnarray}
log_{10}T_{1/2} &=& a \sqrt{\mu} Z_c Z_d Q_\alpha^{-1/2} + b \sqrt{\mu} (Z_c Z_d)^{1/2}+ cI + d I^2 +e [1-(-1)^{l}]  \nonumber \\
 && +fl(l+1)   +g
\label{IUF}
\end{eqnarray}
The values of the parameters for all nuclei are as follows:
$a = 0.3913$,
$b = -1.5172$,
$c = -1.3824$,
$d = 17.9371$,
$e = 0.0341$,
$f = 0.0413$,
$g = -9.9221$.
\item \textbf{MRenB}\\
Modified Ren \textit{et al.} \cite{newrenA2019} formula with the inclusion of nuclear isospin asymmetry and angular momentum shown as:
\begin{equation}
\log_{10} T_{1/2} = a \sqrt{\mu Z_1 Z_2} Q^{-1/2} + b \sqrt{\mu Z_1 Z_2} +  c I + d I^2 + e \left[ l(l+1) \right]+f,
\end{equation}
The parameters a, b, c, d, e, and f are 0.3917, -1.5149, -1.1426, 16.9893, 0.0419, and -10.0276, respectively.
\item \textbf{NMSF}\\
Similarly, Sobiczewaski formula was also modified in the same work \cite{sharma2021npa}:
\begin{eqnarray}
log_{10}T_{1/2}^{NMSF}(sec.) &=& aZ\sqrt{\mu}(Q_{\alpha} - \overline{E}_{i})^{-1/2} + bZ\sqrt{\mu} +  cI + dI^{2} \nonumber\\
 && + el(l+1)+f
\label{nmsf1}
\end{eqnarray}
The parameters a, b, c, d, e, and f are 0.7647, -0.1110, -5.3361, 31.7453, 0.0416, and -29.6290. $\overline{E}_{i}$ is 0 for even-even nuclei, -0.0580 for even-odd, -0.0216 for odd-even and 0.0617 for odd-odd nuclei.
\item \textbf{Royer 2020}\\
In Royer 2020 formula, blocking effect of the unpaired nucleon has been incorporated \cite{royer2020}, which is shown as:
\begin{eqnarray}
log_{10}T_{1/2} = aA^{1/6}\sqrt{Z}+b \frac{Z}{\sqrt{Q_{\alpha}}}+cl(l+1)+d+h
\label{royer2020}
\end{eqnarray}
Adjustable parameters are a=-1.1324, b=1.5966, c=0.0469, d=-25.9523. The value of h for different $\alpha$-decay is 0, -0.2437, -0.0937, and 0.3031 for even-even, even-odd, odd-even, and odd-odd nuclei, respectively.\\
\item \textbf{MTNF}\\
We have modified the Tagepera-Nurmia formula (TNF) formula in our recent work \cite{saxena2021}:
\begin{equation}
\log_{10} T_{1/2}^{\alpha} = a \sqrt{\mu} \left( Z_d Q^{-1/2} - Z_d^{2/3} \right) + bl (l + 1) + c .
\end{equation}
The parameters a, b, and c are  0.8003, 0.0435, and -19.3857, respectively.
\item \textbf{MBudaca}\\
Akrawy \textit{et al.} introduced additional modifications to the UDL for $\alpha$-decay, incorporating the effects of asymmetry and angular momentum. This modified version is referred to as the Modified Budaca Formula (MBudaca) \cite{akrawy2022generalization} and is represented as follows:
\begin{eqnarray}
 log_{10}T_{1/2}&=&a\chi\left(1-d_1\frac{\eta\chi^{2}}{\rho^{2}}\right) + b \rho\left(1-d_2\frac{\eta\chi^{2}}{\rho^{2}}\right)\nonumber\\
 &&+cI+dI^2+e\frac{l(l+1)}{\mu\sqrt{Z_{\alpha} Z_d}(A_{\alpha}^{1/3} +A_d^{1/3})}+f
 \label{eq:mbudac}
\end{eqnarray}
where\\
\begin{equation}
 \chi = Z_{\alpha} Z_d \sqrt{\frac{A_{\alpha} A_d}{(A_{\alpha} +A_d)Q_\alpha}}
\end{equation}

\begin{equation}
 \rho = \sqrt{\frac{Z_{\alpha} Z_d A_{\alpha} A_d (A_{\alpha}^{1/3} +A_d^{1/3})}{(A_{\alpha} +A_d)}}
\end{equation}

\begin{equation}
 \eta =A_{\alpha}^{1/3} +A_d^{1/3}
\end{equation}
\begin{table}[!htbp]
\caption{Coefficients of MBF formula.}
\centering
\def\arraystretch{1.2}
\resizebox{1.0\textwidth}{!}{%
{\begin{tabular}{|c|c|c|c|c|c|c|c|c|}
 \hline
\multicolumn{1}{|c|}{Parity}&
\multicolumn{1}{c|}{a}&
\multicolumn{1}{c|}{d$_1$}&
\multicolumn{1}{c|}{b}&
 \multicolumn{1}{c|}{d$_2$}&
 \multicolumn{1}{c|}{c}&
\multicolumn{1}{c|}{d}&
\multicolumn{1}{c|}{e}&
\multicolumn{1}{c|}{f}\\
 \hline
E-E   & 0.0547  & -0.0005         & 0.0313  & 0.0017           & -8.2558    & -11.9717            & 11.7557    & -20.7548 \\
E-O   & 0.4650 & -0.0213         & -0.1398  & -0.2074           & 42.7512     & -104.0949     & 15.1579     & -33.6109 \\
O-E   & 0.1422 & -0.0427         & -0.0505   & -0.1405           & -5.5379     & 0.5529     & 13.1546    & -19.7348 \\
O-O   & 0.2512 & -0.0253         & -0.0809  & -0.1543           & 5.4532     & -9.1142     & 12.2469  & -23.8240 \\
\hline
\end{tabular}}
}
\label{coefficientMBF}
\end{table}
\item \textbf{QF}\\
We have proposed a new formula i.e. Quadratic fitting formula (QF) \cite{saxena2021}:
\begin{equation}\label{QF1}
log_{10}T_{1/2}^{\alpha}(sec.) = a\sqrt{\mu}\left(\frac{Z_d^{0.6}}{\sqrt{Q_{\alpha}}}\right)^2 + b\sqrt{\mu}\left(\frac{Z_d^{0.6}}{\sqrt{Q_{\alpha}}}\right)+cl(l+1)+d
\end{equation}
The coefficients a, b, c, and d are 0.1890, 2.5713, 0.0424, and -37.9037, respectively.
\item \textbf{NMTN}\\
We have modified Taagepera and Nurmia formula named as new modified Taagepera and Nurmia formula (NMTN) \cite{NMTN}.
\begin{eqnarray}
log_{10}T_{1/2}^{NMTN}(sec.) &=& a \sqrt{\mu \left( Z_d Q_{\alpha}^{-1/2} - Z_d^{2/3} \right)} + b + c \sqrt{l(l+1)} + d \sqrt{I(I+1)} \nonumber \\
 && + h
\label{nmtn}
\end{eqnarray}
\begin{table}[h!]
\centering
\small
\caption{The parameters of NMTN formula for isomeric $\alpha$-decay half-life.}
\begin{tabular}{|c|c|c|}
\hline
\multicolumn{1}{|c}{Parameter}&
\multicolumn{2}{|c|}{Region}\\
\cline{2-3}
\textbf{} & \textbf{$N \leq 126$} & \textbf{$N > 126$} \\ \hline
\textbf{a} & 0.7689 & 0.7054 \\ \hline
\textbf{b} & -19.2355 & -26.4768 \\ \hline
\textbf{c} & 0.3070 & 0.4059 \\ \hline
\textbf{d} & 1.2999 & 19.5486 \\ \hline
\textbf{} & 0.0000 (E-E) & 0.0000 (E-E) \\
\cline{2-3}
\textbf{h} &  -0.1126 (O-A) &  0.0473 (O-A) \\
\cline{2-3}
\textbf{} &  0.2202 (O-O) & -0.1245 (O-O) \\ \hline
\end{tabular}
\label{coeff}
\end{table}
\item \textbf{MUDL}\\
In recognizing the potential contribution of parity effects to the explanation of $\alpha$-decay half-lives, Soylu and Chong have incorporated these effects into their calculations \cite{soylu2021}. Consequently, they have employed the following formula:
\begin{eqnarray}
log_{10}T_{1/2} &=& a Z_c Z_d \sqrt{\frac{\mu}{Q_\alpha}}+ b \sqrt{\mu Z_c Z_d (A_{d}^{1/3} + A_{c}^{1/3})}+ c\sqrt{I(I+1)} \nonumber \\
 &&+d \sqrt{\mu Z_c Z_d (A_{d}^{1/3} + A_{c}^{1/3})} \sqrt{l(l+1)} +e\mu [1-(-1)^l] +f
\label{NUDL}
\end{eqnarray}
The coefficients a, b, c, d, e, and f are 0.3729, -0.4585, 7.6407, 0.0060, -0.0813, and -17.4487, respectively.
\item \textbf{DK1}\\
In 2009, Denisov-Khudenko presented a formula \cite{denisov2010decay} that sheds light on how the half-lives of $\alpha$-decay depend on the angular momentum ($l$) of the $\alpha$ particle.
\begin{eqnarray}
log_{10}T_{1/2} &=& aZQ_{\alpha}^{-1/2}+bA^{1/6}Z^{1/2}\mu ^{-1} +c[l(l+1)]^{1/2}Q_{\alpha} ^{-1}A^{1/6}+d [(-1)^{l}-1]\nonumber \\
 &&+e
\label{DK1}
\end{eqnarray}
\begin{table}[!htbp]
\caption{Coefficients of DK1 formula.}
\centering
\def\arraystretch{0.6}
\resizebox{0.6\textwidth}{!}{%
{\begin{tabular}{|c|c|c|c|c|}
 \hline
\multicolumn{1}{|c|}{Coefficient}&
\multicolumn{1}{c|}{E-E}&
\multicolumn{1}{c|}{E-O}&
\multicolumn{1}{c|}{O-E}&
\multicolumn{1}{c|}{O-O}\\
 \hline
 a&1.4732  &1.4552  &1.4239  & 	1.4925 \\
 b&-1.0459 &-1.1497 &-1.0333 &	-1.0164    \\
 c&1.2557  &2.3101  &1.0811  &	1.0761   \\
 d&0.2376  &1.6223  &0.1613  &-0.0258             \\
 e&-24.0107&-21.1723&-22.7443& -25.0799   \\

  \hline
   \end{tabular}}
   }
   \label{coefficientroyer}
   \end{table}
\item \textbf{Akrawy}\\
Akrawy enhanced his own formulation in 2018 by introducing terms dependent on angular momentum ($l$), as documented in his work \cite{akrawy2018new}. The improved formula is represented as:
 \begin{eqnarray} log_{10}T_{1/2} = aA^{1/6}\sqrt{Z}+b \frac{Z}{\sqrt{Q_{\alpha}}}+c \frac{\sqrt{l(l+1)}}{Q_{\alpha}}+dI+e \mu I^{2}(l(l+1))^{1/4}+f
\end{eqnarray}
The coefficients a, b, c, d, e, and f are -1.3087, 1.3284, 1.9838, 19.5459, 0.5550, and -16.0652, respectively.\\
\item \textbf{ITM}\\
Recently in 2023, the Tavares and Medeiros (TM) formula was improved by us \cite{saxena2023cluster}, which is formulated as:
\begin{eqnarray}
     log_{10}T_{1/2}^{ITM}(sec.) &=& (aZ_{c} + b) \sqrt{\frac{Z_{d}}{Q_{c}}} +(cZ_{c} + d) + e\sqrt{I(I+1)}+ \nonumber\\
 &&f \sqrt{l(l+1)}
    \label{eqmtm1}
    \end{eqnarray}
    The values of the parameters are as follows:
$a = -419.5282$,
$b = 851.4329$,
$c = 0.3070$,
$d = -46.5260$,
$e = 10.2193$,
$f = 0.0263$.
\item \textbf{NMMF}\\
New modified Manjunatha formula \cite{sharma2021npa} is shown as:
\begin{eqnarray}
log_{10}T_{1/2}^{NMMF}(sec.) &=& a\sqrt{\mu}(Z_{d}^{0.4}/\sqrt{Q_{\alpha}})^{2} + b\sqrt{\mu}(Z_{d}^{0.4}/\sqrt{Q_{\alpha}}) + cI + dI^{2}\nonumber\\
 && + el(l+1) +f
\label{nmmf1}
\end{eqnarray}
The parameters a, b, c, d, e, and f are 19.6150, -2.2938, -1.2153, 96.3466, 0.0329, and -65.1623, respectively.
\item \textbf{DK2}\\
 In 2018, Akrawy \textit{et al.} modified Denisov-Khudenko (DK1) formula \cite{akrawy2018}, which represents as:
 \begin{equation}
\log_{10} T_{1/2}^{DK} =   \frac{a Z}{\sqrt{Q}} + \frac{b \sqrt{l(l+1)}}{Q A^{-1/6}} + \frac{c A^{1/6} \sqrt{Z}}{\mu} +d \left( (-1)^l - 1 \right)+e
\end{equation}
The parameters a, b, c, d, and e are 1.4733, -1.0456, 1.2553, 0.2375, and -24.0181, respectively.

\item \textbf{MRF}\\
This is the modified version of RB 2010 formula by Akrawy \textit{et al.} in 2018 \cite{Akrawymrf2018}.
\begin{eqnarray}
log_{10}T_{1/2} &=&  aA^{1/6}\sqrt{Z} + b\frac{Z}{\sqrt{Q_{\alpha}}} +cI+dI^2+ e\frac{ANZ[l(l+1)]^{1/4}}{Q_{\alpha}}+fA[1-(-1)^{l}] \nonumber \\
 &&+g
\label{MRF 2018}
\end{eqnarray}
The values of the parameters are as follows:
$a = -1.4269$,
$b = 1.2387$,
$c = 5.6057$,
$d = 70.1507$,
$e = 0.0001$,
$f = 0.0001$,
$g = -9.9946$.

   \item \textbf{ISEF}\\
   Cheng \textit{et al.} introduced the Improved Semi-Empirical Formula (ISEF) in 2022 \cite{cheng2022isospin}, enhancing it with isospin effects for $\alpha$-decay and cluster radioactivity half-lives. The proposed formula is grounded in the WKB barrier penetrability and written as:
\begin{eqnarray}
log_{10}T_{1/2} = a 2\sqrt{\mu}  Z_d Q_\alpha^{-1/2} + b\sqrt{\mu Z_c Z_d}+c  I_d\sqrt{\mu Z_c Z_d} + dl(l+1)+e
\label{ISEF}
\end{eqnarray}
\begin{table}[!htbp]
\caption{Coefficients of ISEF formula.}
\centering
\def\arraystretch{1.2}
\resizebox{0.6\textwidth}{!}{%
{\begin{tabular}{|c|c|c|c|c|}
 \hline
\multicolumn{1}{|c|}{Coefficient}&
\multicolumn{1}{c|}{E-E}&
\multicolumn{1}{c|}{E-O}&
\multicolumn{1}{c|}{O-E}&
\multicolumn{1}{c|}{O-O}\\
 \hline
 a&0.3028  &0.2583&0.2725&0.3493 \\
 b&-1.8491 &-2.4201   &-1.8630&	 -1.4160   \\
 c&1.0438	&1.9466 &   1.2427&	 0.3418   \\
 d&0.0216  &0.0207      &0.0109&0.0210           \\
 e&6.3531  &23.3042      &9.6506& -7.6359           \\

\hline
\end{tabular}}
}
\label{coefficientISEF}
\end{table}
\end{small}
\end{enumerate}

\newpage

\begin{landscape}
\subsection{Appendix-B}
\begin{table}[!htbp]
\caption{Linear fitting parameters $x$ and $y$ along with the the corresponding $R^2$ value for the 17 empirical formulas shown in Fig.~\ref{linear_fitting}. }
\centering
\def\arraystretch{1.6}
\resizebox{1.54\textwidth}{!}{%
{\begin{tabular}{|c|c|c|c|c|}
 \hline
\multicolumn{1}{|c|}{Sr. No.}&
\multicolumn{1}{c|}{Formula Name}&
\multicolumn{1}{c|}{$x$}&
\multicolumn{1}{c|}{$y$}&
\multicolumn{1}{c|}{$R^2$}\\
 \hline

1 & MYQZR \cite{MYQZR2019} & $a \sqrt{\mu 2 Z_d} Q^{-1/2} + b \sqrt{\mu (2 Z_d)} +  c I + d I^2 +f$ & $ \log_{10}T_{1/2}^{\text{Exp}} - e \frac{l(l+1)}{\mu \sqrt{2 Z_d} A_1^{1/6}}$ & 0.9842 \\
\hline
2 & IUF \cite{ismail2022improved} & $a \sqrt{\mu} Z_c Z_d Q_\alpha^{-1/2} + b \sqrt{\mu} (Z_c Z_d)^{1/2} + cI + d I^2 +g$ & $\log_{10}T_{1/2}^{\text{Exp}} - f l(l+1) -e [1-(-1)^{l}]$ & 0.9835 \\
\hline
3 & MRenB \cite{newrenA2019} & $a \sqrt{\mu Z_1 Z_2} Q^{-1/2} + b \sqrt{\mu Z_1 Z_2} + c I + d I^2 +f$ & $\log_{10}T_{1/2}^{\text{Exp}} - e [l(l+1)]$ & 0.9838 \\
\hline
4 & NMSF \cite{sharma2021npa} & $aZ\sqrt{\mu}(Q_{\alpha} - \overline{E}_{i})^{-1/2} + bZ\sqrt{\mu} + cI + dI^{2} +f$ & $\log_{10}T_{1/2}^{\text{Exp}} - e l (l + 1)$ & 0.9836 \\
\hline
5 & Royer 2020 \cite{royer2020} & $aA^{1/6}\sqrt{Z}+b \frac{Z}{\sqrt{Q_{\alpha}}}+d+h$ & $\log_{10}T_{1/2}^{\text{Exp}} - c l(l+1)$ & 0.9857 \\
\hline
6 & MTNF \cite{saxena2021} & $a \sqrt{\mu} ( Z_d Q^{-1/2} - Z_d^{2/3} ) + c$ & $\log_{10}T_{1/2}^{\text{Exp}} - b l(l+1)$ & 0.9828 \\
\hline
7 & Mbudaca \cite{akrawy2022generalization} & $a\chi\left(1-d_1\frac{\eta \chi^{2}}{\rho^{2}}\right) + b \rho\left(1-d_2\frac{\eta\chi^{2}}{\rho^{2}}\right)+cI+dI^2+f$ & $\log_{10}T_{1/2}^{\text{Exp}}-e\frac{l(l+1)}{\mu\sqrt{Z_{\alpha} Z_d}(A_{\alpha}^{1/3} +A_d^{1/3})}$ & 0.9841 \\
\hline
8 & QF \cite{saxena2021} & $a\sqrt{\mu}\left(\frac{Z_d^{0.6}}{\sqrt{Q_{\alpha}}}\right)^2 + b\sqrt{\mu}\left(\frac{Z_d^{0.6}}{\sqrt{Q_{\alpha}}}\right)+d$ & $\log_{10}T_{1/2}^{\text{Exp}} - c l(l+1)$ & 0.9795 \\
\hline
9 & NMTN \cite{NMTN} & $a\sqrt{\mu}(Z_{d}Q_{\alpha}^{-1/2}-Z_{d}^{2/3})+ d\sqrt{I(I+1)}$ & $\log_{10}T_{1/2}^{\text{Exp}}-b-h-c\sqrt{l(l+1)}$ & 0.9830 \\
\hline
10 & MUDL \cite{soylu2021} & $a Z_c Z_d \sqrt{\frac{\mu}{Q_\alpha}}+ b \sqrt{\mu Z_c Z_d (A_{d}^{1/3} + A_{c}^{1/3})} +c \sqrt{I(I+1)}+ f$ & $\log_{10}T_{1/2}^{\text{Exp}}-d \sqrt{\mu Z_c Z_d (A_{d}^{1/3} + A_{c}^{1/3})} \sqrt{l(l+1)}-e \mu [1-(-1)^l]$ & 0.9837 \\
\hline
11 & DK1 \cite{denisov2010decay} & $aZQ_{\alpha}^{-1/2}+bA^{1/6}Z^{1/2}\mu^{-1} +e$ & $\log_{10}T_{1/2}^{\text{Exp}}-c[l(l+1)]^{1/2}Q_{\alpha}^{-1}A^{1/6} -d[(-1)^{l}-1]$ & 0.9832 \\
\hline
12 & Akrawy \cite{akrawy2018new} & $aA^{1/6}\sqrt{Z}+b \frac{Z}{\sqrt{Q_{\alpha}}}+dI+f$ & $\log_{10}T_{1/2}^{\text{Exp}}-c \frac{\sqrt{l(l+1)}}{Q_{\alpha}}-e \mu I^{2}(l(l+1))^{1/4}$ & 0.98312 \\
\hline
13 & ITM \cite{saxena2023cluster} & $(aZ_{c} + b) \sqrt{\frac{Z_{d}}{Q_{c}}} + (cZ_{c} + d)+e \sqrt{I(I+1)}$ & $\log_{10}T_{1/2}^{\text{Exp}}- f \sqrt{l(l+1)}$ & 0.9697 \\
\hline
14 & NMMF \cite{sharma2021npa} & $a\sqrt{\mu}(Z_{d}^{0.4}/\sqrt{Q_{\alpha}})^{2} + b\sqrt{\mu}(Z_{d}^{0.4}/\sqrt{Q_{\alpha}}) + cI + dI^{2}+f$ & $\log_{10}T_{1/2}^{\text{Exp}}-e l (l + 1)$ & 0.9441 \\
\hline
15 & DK2 \cite{akrawy2018} & $ \frac{a Z}{\sqrt{Q}}+ \frac{c A^{1/6} \sqrt{Z}}{\mu} +e$ & $\log_{10}T_{1/2}^{\text{Exp}}- \frac{b \sqrt{l(l+1)}}{Q A^{-1/6}} - d \left( (-1)^l - 1 \right)$ & 0.9826 \\
\hline
16 & MRF \cite{Akrawymrf2018} & $aA^{1/6}\sqrt{Z} + b\frac{Z}{\sqrt{Q_{\alpha}}} +cI +dI^2+e$ & $\log_{10}T_{1/2}^{\text{Exp}}- e\frac{ANZ[l(l+1)]^{1/4}}{Q_{\alpha}}-fA[1-(-1)^{l}]$ & 0.9848 \\
\hline
17 & ISEF \cite{cheng2022isospin} & $a \sqrt{\mu} 2 Z_d Q_\alpha^{-1/2} + b \sqrt{\mu Z_c Z_d}+c  I_d \sqrt{\mu Z_c Z_d}+e$ & $\log_{10}T_{1/2}^{\text{Exp}}- d l(l+1)$ & 0.9771 \\
\hline
\end{tabular}}
}
\label{coefficientISE}
\end{table}
\end{landscape}
\section{Data availability}
We are committed to supporting open science and data transparency. The authors declare that the experimental data used in the article are available on the National Nuclear Data Center (https://www.nndc.bnl.gov/). The other data supporting the findings of this study are available within the paper in the form of tables. The left datasets used and/or analyzed during the current study are available from the corresponding author upon reasonable request.




\begin{thebibliography}{000}
\bibitem{oganessian2010synthesis} Y. T. Oganessian, F. Sh. Abdullin, P. D. Bailey \textit{et al.}, \textit{Physical review letters} \textbf{104} (2010) 142502.
\bibitem{Og2006} Y. T. Oganessian, V. K. Utyonkov, Yu. V. Lobanov \textit{et al.}, \textit{Physical Review C} \textbf{74} (2006) 044602.
\bibitem{Wuenschel2018} S. Wuenschel, K. Hagel, M. Barbui \textit{et al.}, \textit{Physical Review C} \textbf{97} (2018) 064602.
\bibitem{audi20201} F. Kondev, M Wang, W. J. Huang \textit{et al.}, \textit{Chinese Physics C} \textbf{45} (2021) 030001.
\bibitem{Voinov2020} A. A. Voinov, V. K. Utyonkov, Y. T. Oganessian \textit{et al.}, \textit{Bulletin of the Russian Academy of Sciences: Physics} \textbf{84} (2020) 351.
\bibitem{heinz2012} S. Heinz, \textit{EPJ Web of Conferences} \textbf{38} (2012) 01002.
\bibitem{oganessian2011eleven} Y. T. Oganessian, F. S. Abdullin, P. D. Bailey \textit{et al.}, \textit{Physical Review C} \textbf{83} (2011) 054315.
\bibitem{og2015npa} Y. T. Oganessian and V. Utyonkov, \textit{Nuclear Physics A} \textbf{944} (2015) 62.
\bibitem{gamow1928quantentheorie} G. Gamow, \textit{Zeitschrift fur Physik} \textbf{51} (1928) 204.
\bibitem{hofmann2000discovery} S. Hofmann and G. Munzenberg, \textit{Reviews of Modern Physics} \textbf{72} (2000) 733.
\bibitem{hamilton2013search} J. Hamilton, S. Hofmann, and Y. T. Oganessian, \textit{Annual Review of Nuclear and Particle Science} \textbf{63} (2013) 383.
\bibitem{heenen2015shapes} P.-H. Heenen, J. Skalski, A. Staszczak \textit{et al.}, \textit{Nuclear Physics A} \textbf{944} (2015) 415.
\bibitem{oganessian2015super} Y. T. Oganessian, and V. Utyonkov, \textit{Reports on Progress in Physics} \textbf{78} (2015) 036301.
\bibitem{ni2013} D. Ni, Z. Ren, T. Dong \textit{et al.}, \textit{Physical Review C} \textbf{87} (2013) 024310.
\bibitem{wang2024decay} J. Wang, Z. G. Gan, Z. Y. Zhang \textit{et al.}, \textit{Physics Letters B} \textbf{850} (2024) 138503.
\bibitem{razavy2013quantum} M. Razavy, \textit{World Scientific} (2013).
\bibitem{manjunatha2019} H. C. Manjunatha, L. Seenappa, and K. N. Sridhar, \textit{The European Physical Journal Plus} \textbf{134} (2019) 477.
\bibitem{newrenA2019} D. T. Akrawy,  H. Hassanabadi, Y. Qian \textit{et al.}, \textit{Nuclear Physics A} \textbf{983} (2019) 310.
\bibitem{horoi2004} M. Horoi, \textit{Journal of Physics G: Nuclear and Particle Physics} \textbf{30} (2004) 945.
\bibitem{renA2004} Z. Ren, C. Xu, and Z. Wang, \textit{Physical Review C} \textbf{70} (2004) 034304.
\bibitem{qi2009} C. Qi, \textit{Physical Review Letters} \textbf{103} (2009) 072501.
\bibitem{parkho2005} A. Parkhomenko and A. Sobiczewski, \textit{Acta Physica Polonica B} \textbf{} (2005) 3095.
\bibitem{taageper1961relations} R. Taageper and M. Nurmia, \textit{Ann. Acad. Sci. Fennica Ser. A} \textbf{VI} (1961) 1.
\bibitem{qian2012unfavored} Y. Qian and Z. Ren, \textit{Physical Review C} \textbf{85} (2012) 027306.
\bibitem{singh2020} U. K. Singh, P. K. Sharma, M. Kaushik \textit{et al.}, \textit{Nucl. Phys. A} \textbf{1004} (2020) 122035.
\bibitem{Saxena2021jpg} G. Saxena, P. K. Sharma, and P. Saxena, \textit{Journal of Physics G: Nuclear and Particle Physics} \textbf{48} (2021) 055103.
\bibitem{sharma2021npa} P. K. Sharma, A. Jain, and G. Saxena, \textit{Nuclear Physics A} \textbf{1016} (2021) 122318.
\bibitem{budaca2016} A. Budaca, R. Budaca, and I. Silisteanu, \textit{Nuclear Physics A} \textbf{951} (2016) 60.
\bibitem{poenaru2007} D. N. Poenaru, R. Gherghescu, and N. Carjan, \textit{Europhysics Letters} \textbf{77} (2007) 62001.
\bibitem{royer2010} G. Royer, \textit{Nuclear Physics A} \textbf{848} (2010) 279.
\bibitem{MYQZR2019} D. T. Akrawy and A. H. Ahmed, \textit{Physical Review C} \textbf{100} (2019) 044618.
\bibitem{budaca2021screening} A. Budaca, \textit{The European Physical Journal A} \textbf{57} (2021) 41.
\bibitem{akrawy2022generalization} D. T. Akrawy, A. I. Budaca, G. Saxena \textit{et al.}, \textit{The European Physical Journal A} \textbf{58} (2022) 145.
\bibitem{jain2021nuclear} A. K. Jain, B. Maheshwari, and A. Goel, \textit{Springer} (2021).
\bibitem{garg2023atlas} S. Garg, B. Maheshwari, B. Singh \textit{et al.}, \textit{Atomic data and nuclear data tables} \textbf{150} (2023) 101546.
\bibitem{arnquist2023constraints} I. Arnquist, F. T. Avignone III, A. S. Barabash \textit{et al.}, \textit{Physical Review Letters} \textbf{131} (2023) 152501.
\bibitem{Antalic2016} F. Heßberger, S. Antalic, A. K. Mistry \textit{et al.}, \textit{The European Physical Journal A} \textbf{52} (2016) 192.
\bibitem{nndc} http://www.nndc.bnl.gov/.
\bibitem{Xu2004} F. Xu, E. G. Zhao, R. Wyss \textit{et al.}, \textit{Physical review letters} \textbf{92} (2004) 252501.
\bibitem{Herzberg2006} R.-D. Herzberg, P. T. Greenlees, P. A. Butler \textit{et al.}, \textit{Nature} \textbf{442} (2006) 896.
\bibitem{Chang2007} C. Xu and Z. Ren, \textit{Physical Review C} \textbf{75} (2007) 044301.
\bibitem{Chang2006} C. Xu and Z. Ren, \textit{Nuclear Physics A} \textbf{778} (2006) 1.
\bibitem{Delion2006} D. Delion, S. Peltonen, and J. Suhonen, \textit{Physical Review C} \textbf{73} (2006) 014315.
\bibitem{Peltonen2008} S. Peltonen, D. Delion, and J. Suhonen, \textit{Physical Review C} \textbf{78} (2008) 034608.
\bibitem{Zhangen2008} X. Zhang and Z. Ren, \textit{Journal of Physics G: Nuclear and Particle Physics} \textbf{31} (2005) 959.
\bibitem{Wang2009} Y. Wang, H. F. Zhang, J. M. Dong \textit{et al.}, \textit{Physical Review C} \textbf{79} (2009) 014316.
\bibitem{poenaru1981alpha} D. N. Poenaru and M. Ivascu, \textit{Journal of Physics G: Nuclear Physics} \textbf{7} (1981) 965.
\bibitem{ni2010alpha} D. Ni and Z. Ren, \textit{Journal of Physics G: Nuclear Physics} \textbf{37} (2010) 035104.
\bibitem{santhosh2010alpha} K. P. Santhosh, J. G. Joseph, and S. Sahadevan, \textit{Physical Review C} \textbf{82} (2010) 064605.
\bibitem{qian2011alpha} Y. Qian, Z. Ren, and D. Ni, \textit{Nuclear Physics A} \textbf{866} (2011) 1.
\bibitem{sun2017systematic} X. D. Sun, C. Duan, J. G. Deng \textit{et al.}, \textit{Physical Review C} \textbf{95} (2017) 014319.
\bibitem{sun2017} X. D. Sun, C. Duan, J. G. Deng \textit{et al.}, \textit{Physical Review C} \textbf{95} (2017) 044303.

\bibitem{Denisov2009} V. Y. Denisov and A. Khudenko, \textit{Physical Review C} \textbf{80} (2009) 034603.
\bibitem{Dong2010} J. Dong, H. Zhang, Y. Wang \textit{et al.}, \textit{Nuclear Physics A} \textbf{832} (2010) 198.
\bibitem{clark2024hindrances} R. Clark, \textit{The European Physical Journal Special Topics} (2015) 1.
\bibitem{Anjali2024Systematic} V. Anjali,  K. P. Santhosh, and K. P. Zuhail, \textit{Physica Scripta} \textbf{99} (2024) 055310.
\bibitem{denisov2009alpha} V. Y. Denisov and A. Khudenko, \textit{At. Data Nucl. Data Tables} \textbf{95} (2009) 815.
\bibitem{PhysRevC.98.024302} M. Lewis, \textit{Physical Review C} \textbf{98} (2018) 024302.
\bibitem{fukuchi2005} T. Fukuchi, Y. Gono, A. Odahara \textit{et al.}, \textit{The European Physical Journal A} \textbf{24} (2005) 249.
\bibitem{NMTN} G. Saxena, P. K. Sharma, and P. Saxena, \textit{The European Physical Journal A} \textbf{60} (2024) 50.
\bibitem{royer2020} J.-G. Deng, H. F. Zhang, and G Royer, \textit{Physical Review C} \textbf{101} (2020) 034307.
\bibitem{soylu2021} A. Soylu and C. Qi, \textit{Nuclear Physics A} \textbf{1013} (2021) 122221.
\bibitem{ismail2022improved} M. Ismail, A. Y. Ellithi, A. Adel \textit{et al.}, \textit{The European Physical Journal A} \textbf{58} (2022) 225.
\bibitem{saxena2021} G. Saxena, A. Jain, and P. K. Sharma, \textit{Physica Scripta} \textbf{96} (2021) 125304.
\bibitem{saxena2023cluster} G. Saxena and A. Jain, \textit{The European Physical Journal A} \textbf{59} (2019) 189.

\bibitem{akrawy2018} D. T. Akrawy, H. Hassanabadi, S. S. Hosseini \textit{et al.}, \textit{Nuclear Physics A} \textbf{971} (2018) 130.
\bibitem{akrawy2018new} D. T. Akrawy and A. H. Ahmed, \textit{International Journal of Modern Physics E} \textbf{27} (2018) 1850068.
\bibitem{cheng2022isospin} S. Cheng, W. Wu, L. Cao \textit{et al.}, \textit{The European Physical Journal A} \textbf{58} (2022) 168.
\bibitem{denisov2010decay} V. Y. Denisov, \textit{Physical Review C} \textbf{82} (2010) 059901.


\bibitem{Akrawymrf2018} D. T. Akrawy, H Hassanabadi, S. S. Hosseini \textit{et al.}, \textit{Nuclear Physics A} \textbf{975} (2018) 19.

\bibitem{audi20202} M. Wang,  W. J. Huang, F. G. Kondev \textit{et al.}, \textit{Chinese Physics C} \textbf{45} (2021) 030003.

\bibitem{geiger1911} H. Geiger and J. Nuttall, \textit{The London, Edinburgh, and Dublin Philosophical Magazine and Journal of Science} \textbf{22} (1911) 613.
\bibitem{royer2000} G. Royer, \textit{Journal of Physics G: Nuclear and Particle Physics} \textbf{26} (2000) 1149.

\bibitem{moller2019}  https://t2.lanl.gov/nis/data/astro/molnix96/spidat.html.
\bibitem{seif2023stability}  W. Seif and A. Abdulghany, \textit{Physical Review C} \textbf{108} (2023) 024308.
\bibitem{zhang2018} Y. L. Zhang and Y. Z. Wang, \textit{Physical Review C} \textbf{97} (2018) 014318.
\bibitem{Schuurmans2000} P. Schuurmans, I. Berkes, P. Herzog \textit{et al.}, \textit{Hyperfine Interactions} \textbf{129} (2000) 163.

\bibitem{Severijns2005} N. Severijns, A. A. Belyaev, A. L. Erzinkyan \textit{et al.}, \textit{Physical Review C} \textbf{71} (2005) 044324.
\bibitem{Krause1998} J. Krause, I. Berkes, J. Camps \textit{et al.}, \textit{Physical Review C} \textbf{58} (1998) 3181.

\end{thebibliography}
\end{document}